\documentclass[onecolumn,showpacs,preprintnumbers,amsmath,amssymb]{elsart}
 \input amssym.tex
\newcommand{\hs}{\hspace*{0.5cm}}

\newcommand{\be}{\begin{equation}}
\newcommand{\ee}{\end{equation}}
\newcommand{\bea}{\begin{eqnarray}}
\newcommand{\eea}{\end{eqnarray}}
\newcommand{\nn}{\nonumber}
\newcommand{\crn}{\nonumber \\}

\newcommand{\al}{\alpha}

\newcommand{\bc}{\begin{center}}
\newcommand{\ec}{\end{center}}
\newcommand{\Ga}{\Gamma}

\newcommand {\ba}{\begin{array}}
\newcommand {\ea}{\end{array}}
\newcommand{\ben}{\begin{enumerate}}
\newcommand{\een}{\end{enumerate}}

\usepackage{epsfig, graphicx}
\usepackage{bm}
\usepackage{dcolumn}
\begin{document}
\runauthor{}
\begin{frontmatter}
\title{Supersymmetric reduced minimal 3-3-1 model}
\author[Hanoi]{D. T. Huong\thanksref{H}}
\author[Hanoi]{L. T. Hue\thanksref{H1}}
\author[rio]{ M. C. Rodriguez\thanksref{R}}
\author[Hanoi]{H. N. Long\thanksref{L}}
\address[Hanoi]{Institute of Physics,   Vietnam Academy of Science  and
Technology,
 10 Dao Tan, Ba Dinh, Hanoi, Viet Nam}
\address[rio]{ Funda\c{c}\~{a}o Universidade Federal do Rio Grande-FURG,
Instituto de Matem\'atica, Estat\'\i stica e F\'\i sica-IMEF, Av.
It\'alia, km 8, Campus Carreiros, 96201-900 Rio Grande, RS,
Brazil}

\thanks[H]{Email: dthuong@iop.vast.ac.vn}
\thanks[H1]{Email: lthue@iop.vast.ac.vn}
\thanks[R]{Email: marcosrodriguez@furg.br}
\thanks[L]{Email: hnlong@iop.vast.ac.vn}

\begin{abstract}

We build a  supersymmetric version of the minimal 3-3-1 model with
just two Higgs  triplets using the superfield formalism. We study
the mass spectrum  of all particles in concordance with the
experimental bounds. At the tree level, the masses of charged
gauge bosons are the same as those of charged Higgs bosons. We
also show that the electron, muon and their neutrinos as well as
down and strange quarks gain mass through the loop correction. The
narrow constraint on the ratio $t_w = \frac{w}{w^\prime}$ is given
by studying the new invisible decay mode of the  $Z$ boson.
\end{abstract}
\end{frontmatter}

\section{Introduction}
\label{sec:intro}

Models with $\mathrm{SU(3)}_C\otimes \mathrm{SU(3)}_L\otimes
\mathrm{U(1)}_X$ gauge symmetry (called 3-3-1 models for short)
are interesting possibilities for the physics at the TeV scale
\cite{singer,ppf,Pleitez:1992xh,331rh}. The 3-3-1
 models can have several representation contents depending on the
embedding of the charge operator in the $\mathrm{SU(3)}_{L}$
generators,
\begin{equation}
\frac{Q}{e}= \frac{1}{2}(\lambda_3- \vartheta \lambda_8)+ X\,\ I,
\label{co}
\end{equation}
where the $\vartheta$ parameter defines the different
representation contents, $X$ denotes the $\mathrm{U(1)}_{X}$
charge and $\lambda_3$, $\lambda_8$ are the diagonal generators of
$\mathrm{SU(3)}$.

In fact, this may be the last symmetry involving the lightest
elementary particles: leptons. The lepton sector is exactly the
same as in the Standard Model (SM) \cite{sg} but now there is a
symmetry, at large energies among, say $e^-$, $\nu_e$ and $e^+$.
Once this symmetry is imposed on the lightest generation and
extended to the other leptonic generations it follows that the
quark sector must be enlarged by considering exotic charged
quarks.  It means that some gauge bosons carry lepton and baryon
quantum numbers. Although these models coincide at low energies
with the SM it explains some fundamental questions that are
accommodated, but not explained in the SM, namely
\begin{enumerate}
\item The family number must be three;
\item It explains why $\sin^{2} \theta_{W}<\frac{1}{4}$ is
observed;
\item They are the simplest models that include bileptons
of both types: scalar and vectors ones;
\item It solves the strong CP problem, the Peccei-Quinn
 symmetry occurs also naturally in these models \cite{pal1};
\item The models have several sources of CP
violation \cite{Montero:1998yw,Montero:2005yb};
\item Allow the quantization of electric charge \cite{dongl2};
\item Since one generation of quarks is treated
differently from the others this
may  lead to a natural explanation for  the large mass of the top
quark~\cite{longvan};
\item The models  also  produce a good candidate for
Self-Interacting Dark Matter (SIDM) since there are two Higgs
bosons, one scalar and one pseudoscalar, which have the properties
of candidates for dark matter like stability, neutrality and that
it must not overpopulate the universe~\cite{longlan}, etc.
\end{enumerate}
Another interesting thing about this kind of models is that the
gauge  3-3-1 symmetry is considered a subgroup of the popular $E_6$
Grand Unified Theory (GUT), which can be itself derived from $E_8
\otimes E_8$ heterotic string theory \cite{Meirose:2011cs,ema1}.

In the minimal version, with $\vartheta=\sqrt3$,  the charge
conjugation of the right-handed charged lepton for each generation
is combined with the usual $\mathrm{SU(2)}_L$ doublet of
left-handed leptons components to form an $\mathrm{SU(3)}$ triplet
$(\nu, l, l^c)_L$ \cite{ppf}.  No extra lepton is needed in the
mentioned model, and we shall call such model as minimal 3-3-1
model. There are also another possibility where the triplets
$(\nu, l, L^c)_L$ contain the extra charged leptons  (L). The new
charged leptons (L) do not mix with the known leptons
\cite{Pleitez:1992xh}. We would like to remind that there is no
right-handed (RH) neutrino in both  models. There exists another
interesting possibility ($\vartheta=1/\sqrt3$), where  a
left-handed anti-neutrino to each usual $\mathrm{SU(2)}_L$ doublet
is added to form an $\mathrm{SU(3)}$ triplet $(\nu, l ,
\nu^{c})_L$ \cite{331rh}, and this model is called the 3-3-1 model
with RH neutrinos. The 3-3-1 models have been studied extensively
over the last decade, see for example
\cite{Montero:1998ve,Montero:1999en,Montero:2000ch,Montero:1998sv,Coutinho1,Coutinho3,Coutinho4}.

 Despite attractive properties mentioned above, the usual
3-3-1 models have the weakness  that reduces their predictive
possibility is a plenty  in the scalar sectors. The attempt  to
realize simpler scalar sectors has recently been constructed 3-3-1
model with minimal Higgs sector called the economical 3-3-1 model
\cite{ecn331c,ecn331r}. The 3-3-1 model with minimal content of
fermions and Higgs sector (called the reduced minimal (RM) 3-3-1
model) has also been constructed in   \cite{rm331}.

The supersymmetric version of the  minimal 3-3-1
 model \cite{ppf} has been constructed in
Refs.~\cite{ema1,pal2,331susy1,mcr} (MSUSY331) while the version
with RH neutrinos~\cite{331rh} has already been
constructed in Ref.~\cite{331susy2,huong,Dong:2006vk,Dong:2007qc} (SUSY331RN).
The supersymmetric economical 3-3-1 model  (SUSYE331)
 has been presented recently \cite{Dong:2007qc}.
 Some others
interesting supersymmetric extensions of the 3-3-1 models  were
presented in Ref.
\cite{Sen:2007vx,Sen:2004xt,Diaz:2004fs,Sanchez:2004at,Mira:2003rh}.

In this article we will present a supersymmetric version of the
reduced minimal 3-3-1 model with the triplet $(\nu, l, l^c)_L$
using only two triplets in the scalar sector.

 The outline of the paper is as follows. In
section ~\ref{sec:sm} we present representations of fermions and
Higgs bosons contained in
 the supersymmetric  RM 3-3-1 model.
The super-Lagrangian in terms of superfields is studied in
section.(\ref{lagrangian1}).  In sections
\ref{sec:gaugebosonmass},\ref{sec3},\ref{sec:sp},  we present the
mass eigenstates of gauge bosons, fermions and Higgs bosons as
well as the
 phenomenological consequence of the
 model under consideration. The  Lagrangians in term of fields  are given in
 the Appendix \ref{sec:lagrangian}. In the last section \ref{sec:con},  we
 summary   our results and given conclusions.

\section{The supersymmetric  RM 3-3-1 model}
\label{sec:sm}

In order to consider  supersymmetric model, we first  consider the
particle content in the  model. In this model,  three lepton
superfield families are transformed as the triplet under the
$\mathrm{SU(3)}_{C} \otimes \mathrm{SU(3)}_{L} \otimes
\mathrm{U(1)}_X$ gauge group. We  use  the same notation
for  fermionic field content given in Refs. \cite{331susy1,mcr}
\begin{eqnarray}
 \hat{  L}_{l} &=&
      \left( \begin{array}{c} \hat{ \nu}_{l} \\
                 \hat{ l }\\
                 \hat{ l}^{c}
\end{array} \right)_{L} \sim ({\bf1},{\bf3},0), \,\ l= e, \mu , \tau.
\label{trip}
\end{eqnarray}
In parentheses it appears the transformation properties under the
respective factors
$(\mathrm{SU(3)}_C,~\mathrm{SU(3)}_L,~\mathrm{U(1)}_X)$.

In the quark sector, one quark  superfield
 family is also put
in the triplet representation of $ \mathrm{SU(3)}_L$ as follows
\begin{eqnarray}
  \hat{ Q}_{1L} &=&
      \left( \begin{array}{c} \hat {u}_{1} \\
                 \hat{ d}_{1} \\
                  \hat{J }
\end{array} \right)_{L} \sim \left({\bf3},{\bf3},\frac{2}{3}\right),
\label{q1l}
\end{eqnarray}
and their  respective singlet quark superfields are given by
\begin{equation}
\hat{u}^{c}_{1L} \sim \left({\bf3}^*,{\bf1},-\frac{2}{3}\right),\quad
\hat{d}^{c}_{1L} \sim \left({\bf3}^*,{\bf1},\frac{1}{3}\right),\quad
\hat{J}^{c}_{L} \sim \left({\bf3}^*,{\bf1},-\frac{5}{3}\right),
\label{q1r}
\end{equation}
The remaining two quark  generations are transformed as
 antitriplet superfield representation of $\mathrm{SU(3)}_L$ such as
\begin{equation}
\begin{array}{cc}
\hat{Q}_{2L} =
      \left( \begin{array}{c} \hat{d}_{2} \\
                - \hat{ u}_{2} \\
                 \hat{ j}_{1}          \end{array} \right)_{L},\quad
 \hat{ Q}_{3L} =
      \left( \begin{array}{c} \hat{d}_{3} \\
                - \hat{ u}_{3} \\
                 \hat{ j}_{2}          \end{array} \right)_{L}
\sim \left({\bf3},{\bf3}^{*},-\frac{1}{3}\right) ,
\end{array}
\label{q23l}
\end{equation}
and  their respective singlet superfields are transformed as follows
\begin{eqnarray}
\hat{u}^{c}_{2L} \,\ , \hat{u}^{c}_{3L} &\sim&
\left({\bf3}^*,{\bf1},-\frac{2}{3}\right),\quad \hat{d}^{c}_{2L}
\,\ ,\hat{ d}^{c}_{3L} \sim
\left({\bf3}^*,{\bf1},\frac{1}{3}\right), \quad \nonumber \\
\hat{j}^{c}_{1L} \,\ ,\hat{ j}^{c}_{2L} \,\ &\sim&
\left({\bf3}^*,{\bf1},\frac{4}{3} \right) \,\ . \label{q23r}
\end{eqnarray}
The Eqs.(\ref{q1l},\ref{q23l}) explain exactly the meaning of item
7 given at the introduction of this article.

On the other hand, the scalar superfields which are necessary to generate the
fermion masses are
\begin{equation}
\hat{\rho} =
      \left( \begin{array}{c} \hat{\rho}^{+} \\
                 \hat{ \rho}^{0} \\
                  \hat{\rho}^{++}          \end{array} \right) \sim
({\bf1},{\bf3},+1),\quad
\hat{\chi} =
      \left( \begin{array}{c} \hat{ \chi}^{-} \\
                \hat{  \chi}^{--} \\
                \hat{  \chi}^{0}          \end{array} \right) \sim
({\bf1},{\bf3},-1).
\label{3t}
\end{equation}

To remove chiral anomalies generated by the
superpartners of the scalars,  we have to introduce two other scalar superfields as follows
\begin{equation}
\hat{\rho}^{\prime} =
      \left( \begin{array}{c} \hat{\rho}^{\prime-} \\
                  \hat{\rho}^{\prime0} \\
                  \hat{\rho}^{\prime--}          \end{array} \right)_L
\sim ({\bf1},{\bf3}^{*},-1),\quad
\hat{\chi}^{\prime} =
      \left( \begin{array}{c} \hat{\chi}^{\prime+} \\
                  \hat{\chi}^{\prime++} \\
                  \hat{\chi}^{\prime0}          \end{array} \right)_L
\sim ({\bf1},{\bf3}^{*},+1).
\label{shtc}
\end{equation}
 It is to be  noted  that the superfields formalism is useful in writing the Lagrangian which is
manifestly invariant under the supersymmetric
transformations~\cite{wb} with fermions and scalars put in chiral
superfields while the gauge bosons in vector superfields. As
usual,  the superfield of a field $\phi$ will be denoted by
 $\hat{\phi}$~\cite{haber}. The chiral superfield of a
multiplet $\phi$ is denoted by
\begin{eqnarray}
\hat{\phi}(x,\theta,\bar{\theta})&=& \tilde{\phi}(x) + i \; \theta
\sigma^{\mu} \bar{ \theta} \; \partial_{\mu} \tilde{\phi}(x)
+\frac{1}{4} \; \theta \theta \; \bar{ \theta}\bar{ \theta} \;
\square
\tilde{\phi}(x) \nonumber \\
& & \mbox{} +  \sqrt{2} \; \theta \phi(x) + \frac{i}{ \sqrt{2}} \;
\theta \theta \; \bar{ \theta} \bar{ \sigma}^{\mu}
\partial_{\mu}\phi(x)
\nonumber \\ && \mbox{}+  \theta \theta \; F_{\phi}(x).
\label{phi}
\end{eqnarray}

Concerning the gauge bosons and their superpartners, if we denote
the gluons by $g^b$ the respective superparticles, the gluinos,
are denoted by $\lambda^b_{C}$, with $b=1, \ldots,8$; and in the
electroweak sector we have $V^b$, the gauge boson of
$\mathrm{SU(3)}_{L}$, and their gaugino partners $\lambda^b_{A}$;
finally we have the gauge boson of $U(1)_{X}$, denoted by
$\hat{B}$, and its supersymmetric partner $\lambda_{B}$.

The vector superfield is given by
\begin{eqnarray}
\hat{V}(x,\theta,\bar\theta)&=&-\theta\sigma^\mu\bar\theta
V_{\mu}(x) +i\theta\theta\bar\theta
\overline{\lambda}(x)-i\bar\theta\bar\theta\theta
\lambda(x)\nonumber \\
&+&\frac{1}{2}\theta\theta\bar\theta\bar\theta D(x).
\label{vector}
\end{eqnarray}

As the other version of the  $\mathrm{SU(3)}_c \otimes
\mathrm{SU(3)}_L \otimes \mathrm{U(1)}_X$, the vector superfields
for the gauge bosons of each factor $\mathrm{SU(3)}_C$,
$\mathrm{SU(3)}_L$ and $\mathrm{U(1)}_X$ are denoted by
$\hat{V}_C,\hat{\bar{V}}_C$; $\hat{V},\hat{\bar{V}}$; and
$\hat{V^\prime}$, respectively, where we have defined
\bea \hat{V}_{C}&=&T^{a}\hat{V}^{a}_{C}, \,\
\hat{\bar{V}}_{C}=\bar{T}^{a}\hat{V^{a}}_{C}, \,\ a=1,\cdots,8;
\crn \hat{V}&=&T^{a}\hat{V}^{a}, \,\
\hat{\bar{V}}=\bar{T}^{a}\hat{V^{a}},\crn \hat{V^\prime}&=& T^9
\hat{B}, \label{bosoncouplfer} \eea where $T^a=\lambda^a/2$,
$\bar{T}^a=-\lambda^{*a}/2$ are the generators of triplet and
antitriplet  representations, respectively, and $\lambda^a$ are
the Gell-Mann matrices, and the $T^9= (1/\sqrt{6})\,
\mathrm{diag}(1,~1,~1)$ is the generator of $\mathrm{U(1)}_X$
which satisfies  the relation: $\mathrm{Tr}(T^aT^b)= 1/2
\delta_{ab}$ with all $a,b=1,2,..9$.

\section{The Lagrangian}
\label{lagrangian1}

With the superfields introduced in the last section we can  build  an  invariant
supersymmetric  Lagrangian.   As  usual as  in supersymmetric model,  for the model under
consideration, we have
\begin{equation}
  \mathcal{L}_{3-3-1} =\mathcal{L}_{\mathrm{SUSY}} +\mathcal{L}_{\mathrm{soft}}.
\label{l1}
\end{equation}
Here $\mathcal{L}_{\mathrm{SUSY}}$ is the supersymmetric piece,
while $\mathcal{L}_{\mathrm{soft}}$ explicitly breaks
supersymmetry. Below we will write each of these Lagrangians in
terms of the respective superfields.

\subsection{The supersymmetric terms}
\label{subsec:st}

The supersymmetric terms can be divided as follows
\begin{equation}
  \mathcal{L}_{\mathrm{SUSY}} =  \mathcal{L}_{\mathrm{Lepton}}
                  +\mathcal{L}_{\mathrm{Quarks}}
                  +\mathcal{L}_{\mathrm{Gauge}}
                  +\mathcal{L}_{\mathrm{Scalar}},
\label{l2}
\end{equation}
where each term is given by
\begin{equation}
 \mathcal{L}_{\mathrm{Lepton}}
     = \int d^{4}\theta\;\left[\,\hat{ \bar{L}}e^{2g\hat{V}}
 \hat{L} \,\right],
\label{l3}
\end{equation}
\begin{eqnarray}
\mathcal{L}_{\mathrm{Quarks}}
     &=& \int d^{4}\theta\;\left[\,\hat{ \bar{Q}}_{1}
e^{2[g_{s}\hat{V}_{c}+g\hat{V}+(2g'/3)\hat{V}']} \hat{Q}_{1}
+\,\hat{\bar{Q}}_{\alpha}
e^{2[g_{s}\hat{V}_{c}+g\hat{\bar{V}}-(g'/3)\hat{V}']} \hat{Q}_{\alpha} \,\right.
\nonumber \\&+&
\left.\,\hat{ \bar{u}}_{i}
e^{2[g_{s}\hat{\bar{V}}_{c}-(2g'/3)\hat{V}']} \hat{u}_{i}
+\hat{ \bar{d}}_{i}
e^{2[g_{s}\hat{\bar{V}}_{c}+(g'/3)\hat{V}']} \hat{d}_{i}\right.
\nonumber \\ &+&\left. \,\hat{ \bar{J}}
e^{2[g_{s}\hat{\bar{V}}_{c} -(5g'/3)\hat{V}']} \hat{J}
+ \hat{ \bar{j}}_i
e^{2[g_{s}\hat{\bar{V}}_{c} +(4g'/3)\hat{V}']} \hat{j}_i\right]
\label{l4}
\end{eqnarray}
where the sum for $i=1,2,3$, $\alpha =1,2$ and
\begin{eqnarray}
 \mathcal{L}_{\mathrm{Gauge}} = \frac{1}{4}&\times&\left[ \int  d^{2}\theta\;
         \left( {W}^{a}_{c}{W}^{a}_{c}+{W}^{a}{W}^{a}+{W}^{ \prime}{W}^{
         \prime}\right)\right.\nonumber\\&+&\left.
       \int  d^{2}\bar{\theta}\;
         \left( \bar{W}^{a}_{c}\bar{W}^{a}_{c}+\bar{W}^{a}\bar{W}^{a}+
 \bar{W}^{ \prime}\bar{W}^{ \prime}\right)\,\right],
\label{l5}
\end{eqnarray}
where $\hat{V}_{c},\hat{\bar{V}}_{c}$, $\hat{V}$ and
$\hat{\bar{V}}$ are defined in  Eq.(\ref{bosoncouplfer}) and
$g_{s},g$ and $g^{\prime}$ are the gauge couplings of
$\mathrm{SU(3)}_{C},~\mathrm{SU(3)}_L$ and $\mathrm{U(1)}_X$,
respectively. $W^{a}_{c}$, $W^{a}$ and $W^{ \prime}$ are the
strength fields, and they are given by \bea W^{a}_{\alpha c}&=&-
\frac{1}{8g_{s}} \bar{D} \bar{D} e^{-2g_{s} \hat{V}_{c}}
D_{\alpha} e^{-2g_{s} \hat{V}_{c}}, \crn W^{a}_{\alpha}&=&-
\frac{1}{8g} \bar{D} \bar{D} e^{-2g \hat{V}} D_{\alpha} e^{-2g
\hat{V}}, \crn W^{\prime}_{\alpha}&=&- \frac{1}{4} \bar{D} \bar{D}
D_{\alpha} \hat{V}^{\prime} \,\ . \label{l6} \eea Finally, the
Lagrangian for the Higgs superfield  is given as follows
\begin{eqnarray}
 \mathcal{L}_{\mathrm{Scalar}}
     &=&  \int d^{4}\theta\;\left[\,\hat{ \bar{ \rho}}e^{2g\hat{V}+g'\hat{V}'}
\hat{ \rho} +
\hat{ \bar{ \chi}}e^{2g\hat{V}-g'\hat{V}'}
\hat{ \chi} + \hat{\bar{ \rho}}^\prime e^{2g\hat{\bar V}-g'\hat{V}'}
\hat{ \rho}^\prime + \hat{ \bar{ \chi}}^\prime e^{2g\hat{\bar V}+g'\hat{V}'}
\hat{ \chi}^\prime \right]
\nonumber \\ &+&
\int d^2\theta\, W+\int d^2\bar\theta\, \bar{W},
\label{l7}
\end{eqnarray}
where $W$ is the superpotential that is
written details in the next subsection.  After integrating the super-Lagrangian
given in Eqs.(\ref{l3},\ref{l4},\ref{l5}) and Eq.(\ref{l7}),
we  obtain  the Lagrangian
given in Appendix \ref{sec:lagrangian}.

\subsection{Superpotential.}
\label{subsec:spotential}
 Let us write the full  superpotential in the  model under consideration.
The superpotential which is invariant under $\mathrm{SU(3)}_C
\otimes \mathrm{SU(3)}_L \otimes \mathrm{U(1)}_X$ group can be
written by
\begin{equation}
W=\frac{W_{2}}{2}+ \frac{W_{3}}{3},
\label{sp1}
\end{equation}
with $W_{2}$ is a combination of two chiral
superfields  and the terms permitted by
the considered  symmetry are
\begin{equation}
W_{2}=\mu_{ \rho} \hat{ \rho} \hat{ \rho}^\prime+
\mu_{ \chi} \hat{ \chi} \hat{ \chi}^\prime,
\label{sp2}
\end{equation}
and  $W_{3}$ is invariant under the mentioned symmetry and a combination of
three chiral superfields. That term has the  following form
\begin{eqnarray}
W_{3}&=& \sum_{a,b,c}\lambda_{1abc} \epsilon \hat{L}_{a} \hat{L}_{b} \hat{L}_{c}+
\sum_{a}\lambda_{2a}\epsilon\hat{L}_{a}\hat{\chi}\hat{\rho}+\sum_i
\kappa_{1i} \hat{Q}_{1} \hat{\rho}^{\prime} \hat{d}^{c}_{i} +
\kappa_{2} \hat{Q}_{1} \hat{\chi}^\prime \hat{J}^{c}\crn  &+&
\sum_{\alpha i} \kappa_{3\alpha i} \hat{Q}_{\alpha} \hat{\rho} \hat{u}^{c}_{i}
+ \sum_{\alpha \beta}
\kappa_{4\alpha\beta}\hat{Q}_{\alpha } \hat{\chi} \hat{j}^{c}_{\beta}
+ \sum_{\alpha ij}\kappa_{5\alpha ij} \hat{Q}_{\alpha} \hat{L}_{i}
\hat{d}^{c}_{j} \crn &+&
\sum_{i,j,k}\xi_{1ijk} \hat{d}^{c}_{i} \hat{d}^{c}_{j} \hat{u}^{c}_{k}
+
\sum_{ij\beta}
\xi_{2ij\beta} \hat{u}^{c}_{i} \hat{u}^{c}_{j} \hat{j}^{c}_{\beta}+
\sum_{i\beta}
\xi_{3 i\beta} \hat{d}^{c}_{i} \hat{J}^{c} \hat{j}^{c}_{\beta},
\label{sp3}
\end{eqnarray}
with $i,j,k=1,2,3$, $\alpha=2,3$ and $\beta=1,2$. The terms $\kappa_{5}$
and $\xi_{2}$ will induce the proton decay as shown at \cite{pal2}.

Choosing, as we have done in \cite{massspectrum}, the following R-charges
\begin{eqnarray}
n_{\rho^{\prime}}&=&-1, \,\
n_{\rho}=1, \,\
n_{\chi}=n_{\chi^{\prime}}=0, \nonumber \\
n_{L}&=&n_{Q_{i}}=n_{d_{i}}=1/2, \,\
n_{J_{i}}=-1/2, \,\ n_{u}=-3/2,
\label{rdiscsusy331}
\end{eqnarray}
it is easy to see that all the fields  $\chi$, $\chi^{\prime}$,
$\rho$, $\rho^{\prime}$, $L$, $Q_{i}$, $u$, $d$ and $J_{i}$ have
R-charge equal to one, while their superpartners have opposite
R-charge. This kind of symmetry is similar to that in the MSSM.
The superpotential which satisfies the  R- symmetry given in
(\ref{rdiscsusy331}) can be written by
\begin{eqnarray}
W&=&\frac{\mu_{ \rho}}{2} \hat{ \rho} \hat{ \rho}^\prime+\frac{
\mu_{ \chi}}{2} \hat{ \chi} \hat{ \chi}^\prime + \frac{1}{3} \left
[ \sum_{a,b,c}\lambda_{1abc} \epsilon \hat{L}_{a} \hat{L}_{b}
\hat{L}_{c}\right.\crn& +&\left.
\sum_{a}\lambda_{2a}\epsilon\hat{L}_{a}\hat{\chi}\hat{\rho}+\sum_i
\kappa_{1i} \hat{Q}_{1} \hat{\rho}^{\prime} \hat{d}^{c}_{i}+
\kappa_{2} \hat{Q}_{1} \hat{\chi}^\prime \hat{J}^{c} \right.  \nonumber \\
&+& \left. \sum_{\alpha i} \kappa_{3\alpha i} \hat{Q}_{\alpha} \hat{\rho}
\hat{u}^{c}_{i} + \sum_{\alpha \beta}
\kappa_{4\alpha\beta}\hat{Q}_{\alpha } \hat{\chi}
\hat{j}^{c}_{\beta} + \sum_{\alpha ij}\kappa_{5\alpha ij}
\hat{Q}_{\alpha} \hat{L}_{i} \hat{d}^{c}_{j} \right ]
\label{sp1}
\end{eqnarray}
Based on the  superpotential given  in Eq.(\ref{sp1}), we can
generate mass to neutrinos and recover all the nice consequences
given in \cite{massspectrum}. We will consider these
 details in the next section.

\subsection{Broken structure from SUSY RM 3-3-1 to
$\mathrm{SU(3)}_{C} \otimes \mathrm{U(1)}_{Q}$.}
\label{breaksusy331}

The pattern of the symmetry breaking of the model is given by the
following scheme (using the notation given in \cite{massspectrum})
\bea \mbox{SUSY RM
3-3-1}&\stackrel{\mathcal{L}_{\mathrm{soft}}}{\longmapsto}&
\mbox{SU(3)}_C\ \otimes \ \mbox{SU(3)}_{L}\otimes
\mbox{U(1)}_{X}\crn&\stackrel{\langle\chi\rangle \langle
\chi^{\prime}\rangle}{\longmapsto}& \mbox{SU(3)}_{C} \ \otimes \
\mbox{SU(2)}_{L}\otimes \mbox{U(1)}_{Y} \crn
&\stackrel{\langle\rho\rangle \langle
\rho^{\prime}\rangle}{\longmapsto}& \mbox{SU(3)}_{C} \ \otimes \
\mbox{U(1)}_{Q}. \label{breaksusy331tou1} \eea For the sake of
simplicity, here we assume that vacuum expectation values (VEVs)
are real. This means that the CP violation through the scalar
exchange is not considered in this work.  Note that
non-supersymmetric 3-3-1 model with non-real VEV was studied at
\cite{Montero:1998yw,Montero:2005yb} and it is the point 5 given
in the introduction.

When one breaks the 3-3-1 symmetry to the $\mathrm{SU(3)}_{C}
\otimes \mathrm{U(1)}_{Q}$, the scalar fields get the following
VEVs:
\begin{eqnarray}
< \rho > &=&
      \left( \begin{array}{c} 0 \\
                  u \\
                  0          \end{array} \right),\quad
< \chi > =
      \left( \begin{array}{c} 0 \\
                  0 \\
                  w          \end{array} \right), \crn
< \rho^{\prime} > &=&
      \left( \begin{array}{c} 0 \\
                  u^{\prime} \\
                  0          \end{array} \right),\quad
< \chi^{\prime} > =
      \left( \begin{array}{c} 0 \\
                  0 \\
                  w^{\prime}          \end{array} \right),
\label{vev1}
\end{eqnarray}
where $u=v_{\rho}/ \sqrt{2}$,
$w=v_{\chi}/ \sqrt{2}$, $u^{\prime}=v_{\rho^{\prime}}/ \sqrt{2}$ and
$w^{\prime}=v_{\chi^{\prime}}/ \sqrt{2}$.
Because of the  pattern of the symmetry breaking given in (\ref{breaksusy331tou1}),
 the VEVs of the model under consideration  have to be satisfied the conditions:
\begin{equation}
w,w^{\prime} \gg u,u{^\prime}.\label{cond}
\end{equation}
On the other hand, the constraint on the $W$ bosons mass \cite{mcr},  see
Eq.(\ref{wmass}), we get the following constraint on
$V^{2}_{\rho}$
\begin{equation}
V^{2}_{\rho}=(246\;{\rm GeV})^2
\label{wmasslimite}
\end{equation}
where $V^{2}_{\rho}= v^{2}_{\rho}+v^{\prime 2}_{\rho}$.

\subsection{Soft terms}

The most general soft supersymmetry breaking terms,
which do not induce quadratic divergence, are described by
Girardello and Grisaru \cite{10}. They found that the allowed
terms can be categorized as follows:
\begin{itemize}
\item The scalar mass term
\begin{equation}
\mathcal{L}_{\mathrm{SMT}}=-m^{2} A^{\dagger}A,
\end{equation}
\item The  gaugino mass  term
\begin{equation}
\mathcal{L}_{\mathrm{GMT}}=- \frac{1}{2} (M_{ \lambda} \lambda^{a}
\lambda^{a}+\mathrm{H.c.})
\end{equation}
\item The scalar interaction terms
\end{itemize}
\begin{equation}
\mathcal{L}_{\mathrm{int}}= m_{ij}A_{i}A_{j}+
f_{ijk}\epsilon^{ijk}A_{i}A_{j}A_{k}+\mathrm{H.c.}
\end{equation}
The soft SUSY breaking parameters are in general complex and they also can generate
SUSY flavor problem. Therefore we can expect that in this model,   there are several sources
of CP violation as well as flavor problem.  This subject can be explored in the future.

In the model, the soft terms must be consistent with the
 3-3-1 gauge symmetry. Hence, the soft terms have
the following form
\begin{equation}
\mathcal{L}_{\mathrm{soft}}=\mathcal{L}_{\mathrm{GMT}}
+\mathcal{L}_{\mathrm{SMT}}+\mathcal{L}_{\mathrm{int}},
\end{equation}
where
\begin{eqnarray}
\mathcal{L}_{\mathrm{GMT}}=- \frac{1}{2} \left[m_{ \lambda_{C}}
\sum_{a=1}^{8} \left( \lambda^{a}_{C} \lambda^{a}_{C} \right) +m_{
\lambda} \sum_{a=1}^{8} \left( \lambda^{a}_{A} \lambda^{a}_{A}
\right) +m^{ \prime} \lambda_{B} \lambda_{B}+\mathrm{H.c.}
\right], \label{gmt1}
\end{eqnarray}
where $\lambda_{C}$ are the gluinos, $\lambda_{A}$ are the
gauginos of $\mathrm{SU(3)}$ and $\lambda_{B}$ is the gauginos of
$\mathrm{U(1)}$ [see Eq.(\ref{tutty})]. The gauginos get their
masses at SUSY broken scale while their superpartners (the gauge
bosons) are massless at this scale, because  their masses  appear
only after we break the symmetry $\mathrm{SU(3)}_{L}\otimes
\mathrm{U(1)}_{X}$ [ see Eq.(\ref{wmass})] in the next section.
The second term which gains masses to all the scalars is written
as
\begin{eqnarray}
\mathcal{L}_{\mathrm{SMT}}&=& -m^2_{ \rho}\rho^{ \dagger}\rho-
m^2_{ \chi}\chi^{ \dagger}\chi -m^2_{\rho^{\prime}}\rho^{\prime
\dagger}\rho^{\prime}- m^2_{\chi^{\prime}}\chi^{\prime
\dagger}\chi^{\prime} \nonumber\\&-&m_{L}^{2}
\hat{L}^{\dagger}_{aL} \hat{L}_{aL}- m_{Q_{\alpha}}^{2}
\hat{Q}^{\dagger}_{\alpha L} \hat{Q}_{\alpha L} -m_{Q_3}^{2}
\hat{Q}^{\dagger}_{3L} \hat{Q}_{3L}\nonumber \\ &-& m_{u_{i}}^2
\hat{u}^{c \dagger}_{iL} \hat{u}^{c}_{iL}- m_{d_{i}}^2 \hat{d}^{c
\dagger}_{iL} \hat{d}^{c}_{iL}- m_{J}^{2} \hat{J}^{c \dagger}_{L}
\hat{J}^{c}_{L} - m_{j_{ \beta}}^{2} \hat{j}^{c \dagger}_{ \beta
L} \hat{j}^{c}_{ \beta L} \label{smt}
\end{eqnarray}
and the last term is given by
\begin{eqnarray}
\mathcal{L}_{\mathrm{int}}&=& \left[ \varepsilon_{0abc}  \epsilon
\hat{L}_{aL} \hat{L}_{bL} \hat{L}_{cL}+ \varepsilon_{1ab} \epsilon
\hat{L}_{aL} \chi \rho  + \hat{Q}_{\alpha L} \left(
\omega_{1\alpha i} \rho \hat{u}^{c}_{iL}+ \omega_{3 \alpha aj}
\hat{L}_{aL} \hat{d}^{c}_{jL} \right. \right. \crn&+& \left.
\left. \omega_{4\alpha \beta}  \chi \hat{j}^{c}_{\beta L} \right)
+ \hat{Q}_{3L}( \zeta_{1i}  \rho^{\prime} \hat{d}^{c}_{iL}+
\zeta_{3J} \chi^{\prime} \hat{J}^{c}_{L}) + \varsigma_{1ijk}
\hat{d}^{c}_{iL} \hat{d}^{c}_{jL} \hat{u}^{c}_{kL}\right. \crn &+&
\left. \varsigma_{2i\beta} \hat{d}^{c}_{iL} \hat{J}^{c}_{L}
\hat{j}^{c}_{\beta L} +\varsigma_{3ij\beta} \hat{u}^{c}_{iL}
\hat{u}^{c}_{jL} \hat{j}^{c}_{\beta L}+\mathrm{H.c.} \right].
\end{eqnarray}

\section{Gauge boson masses}
\label{sec:gaugebosonmass}
Just as it did
in the usual 3-3-1 model \cite{ppf,mcr,massspectrum}, we can
 divide the gauge boson masses into two parts namely
the charged and neutral gauge boson masses. The mass Lagrangian for the gauge
bosons can be obtained by
\begin{eqnarray}
\mathcal{L}^{\mathrm{gauge}}_{\mathrm{mass}}&=&
 \left(
 \begin{array}{ccc}
   0 &
   0 &
   \frac{v_{\chi}}{\sqrt{2}} \\
 \end{array}
 \right)
 \left ( \frac{g}{2}\lambda^a
 V_a^{\mu}-\frac{g^\prime}{\sqrt{6}}B^{\mu}\right)^2
\left(
 \begin{array}{ccc} 0 & 0 & \frac{v_{\chi}}{\sqrt{2}} \\ \end{array}
 \right)^T   \crn &+&
  \left(
  \begin{array}{ccc}
    0 &
    0 &
    \frac{v_{\chi^\prime}}{\sqrt{2}} \\
  \end{array}
  \right)
  \left ( -\frac{g}{2}\lambda^{*a}
  V_a^{\mu}+\frac{g^\prime}{\sqrt{6}}B^{\mu}\right)^2
 \left(
  \begin{array}{ccc} 0 & 0 & \frac{v_{\chi^\prime}}{\sqrt{2}} \\ \end{array}
\ \right)^T  \crn
 &+&
  \left(
  \begin{array}{ccc}
    0 &
    \frac{v_{\rho}}{\sqrt{2}}  &
    0 \\
  \end{array}
  \right)
  \left ( \frac{g}{2}\lambda^{a}
  V_a^{\mu}+\frac{g^\prime}{\sqrt{6}}B^{\mu}\right)^2
 \left(
  \begin{array}{ccc} 0 &  \frac{v_{\rho}}{\sqrt{2}} & 0 \\ \end{array}
\ \right)^T   \crn &+&
  \left(
  \begin{array}{ccc}
    0 &
    \frac{v_{\rho^\prime}}{\sqrt{2}}  &
    0 \\
  \end{array}
  \right)
  \left ( -\frac{g}{2}\lambda^{*a}
  V_a^{\mu}-\frac{g^\prime}{\sqrt{6}}B^{\mu}\right)^2
 \left(
  \begin{array}{ccc} 0 &  \frac{v_{\rho^\prime}}{\sqrt{2}} & 0 \\ \end{array}
\ \right)^T. \label{euq1}\end{eqnarray}
The Lagrangian  in Eq.(\ref{euq1}) produces the charged gauge
boson mass terms given as follows
\begin{eqnarray}
\mathcal{L}_{\mathrm{mass}}^{\mathrm{charged}} &=& M^{2}_{W}
W^{-}_{\mu}W^{+ \mu}+ M^{2}_{V} V^{-}_{\mu} V^{+ \mu}+ M^{2}_{U}
U^{--}_{\mu}U^{++ \mu}, \label{bgmcc}
\end{eqnarray}
with
\begin{eqnarray}
M^{2}_{U}&=& \frac{g^{2}}{8}(v^{2}_{ \rho}+v^{2}_{ \chi}+
v^{2}_{ \rho^{\prime}}+v^{2}_{ \chi^{\prime}}),
\nonumber \\
M^{2}_{W}&=& \frac{g^{2}}{8}(v^{2}_{ \rho}+v^{2}_{ \rho^{\prime}}), \nonumber \\
M^2_{V}&=& \frac{g^{2}}{8}(v^{2}_{ \chi}+v^{2}_{ \chi^{\prime}})
\label{wmass}
\end{eqnarray}
 and the mass eigenvectors are given respectively
 \begin{eqnarray}
 W^{ \pm}_{\mu}(x)&=&\frac{1}{\sqrt{2}}\left[V^{1}_{\mu}(x) \mp i V^{2}_{\mu}(x)\right],
 \crn
 V^{ \pm}_{\mu}(x)&=&\frac{1}{\sqrt{2}}\left[V^{4}_{\mu}(x) \pm i V^{5}_{\mu}(x)\right],
 \nonumber \\
 U^{\pm \pm}_{\mu}(x) &=&\frac{1}{\sqrt{2}}\left[V^{6}_{\mu}(x) \pm i
 V^{7}_{\mu}(x)\right].
 \label{defcarbosons}
 \end{eqnarray}
 Before continuing  we  note
\begin{equation}
M^{2}_{W}+M^{2}_{V}= \frac{g^{2}}{8} \left( v^{2}_{ \rho}+v^{2}_{ \rho^{\prime}}+
v^{2}_{ \chi}+v^{2}_{ \chi^{\prime}}\right).
\label{comparreferre}
\end{equation}

The neutral gauge bosons $(V_{3}^{\mu},V_{8}^{\mu},B^{\mu})$ are
mixing. The  mass Lagrangian for neutral gauge bosons is given as
\begin{eqnarray}
\mathcal{L}_{\mathrm{mass}}^{\mathrm{neutral}}=
 \left(
 \begin{array}{ccc}
   V_{3}^{\mu} & V_{8}^{\mu} & B^{\mu} \\
 \end{array}
 \right)
 M_{NG}^2
 \left(
 \begin{array}{ccc}
   V_{3 \mu} & V_{8 \mu} & B_{\mu} \\
 \end{array}
 \right)^T
\end{eqnarray}
with
\begin{eqnarray}
\scriptsize{M_{\mathrm{NG}}^2= \frac{g^2}{4}\left(
\begin{array}{ccc}
  \frac{v^2_\rho+v^2_{\rho^\prime}}{2} &
  -\frac{v^2_\rho+v^2_{\rho^\prime}}{\sqrt{3}}
  & -t\sqrt{\frac{2}{3}}(v^2_\rho+v^2_{\rho^\prime}) \\
  -\frac{v^2_\rho+v^2_{\rho^\prime}}{\sqrt{3}} & \frac{1}{3}
  \left(v^2_\rho+v^2_{\rho^\prime}+4v_\chi^2+4 v^2_{\chi^\prime} \right)
  & \frac{2t}{3\sqrt{2}}(v_\rho^2+v_{\rho^\prime}^2+2v_\chi^2+2v^2_{\chi^\prime}) \\
  -t\sqrt{\frac{2}{3}}(v^2_\rho+v^2_{\rho^\prime}) & \frac{2t}{3\sqrt{2}}
  (v_\rho^2+v_{\rho^\prime}^2+2v_\chi^2+2v^2_{\chi^\prime})
   & \frac{2t^2}{3}(v_\rho^2+v_{\rho^\prime}^2+v_\chi^2+v_{\chi^\prime}^2) \\
\end{array}
\right)}.\crn \label{mass1}\end{eqnarray} After diagonalization
the matrix $M^2_{\mathrm{NG}}$, we obtain the mass eigenvalues as
follows
\begin{eqnarray}
 M^2_\gamma &=&0,  \crn
 M^2_{Z} &=& \frac{g^2(2+t^2)}{24}\left(v_{\rho}^2+v_{\rho^\prime}^2
+v_\chi^2+v_{\chi^\prime}^2\right.\nonumber\\
&&\left.-\sqrt{\frac{-4(3+2t^2)}{2+t^2}
 (v_\rho^2+v^2_{\rho^\prime})(v_\chi^2+v_{\chi^\prime}^2)+
 (v_\rho^2+v_{\rho^\prime}^2+v_\chi^2+v^2_{\chi^\prime})^2}
 \right),
 \crn
 M^2_{Z^\prime} &=& \frac{g^2(2+t^2)}{24}\left(v_{\rho}^2+v_{\rho^\prime}^2
 +v_\chi^2+v_{\chi^\prime}^2 \right.\nonumber\\  &&\left.+\sqrt{\frac{-4(3+2t^2)}{2+t^2}
 (v_\rho^2+v^2_{\rho^\prime})(v_\chi^2+v_{\chi^\prime}^2)+
 (v_\rho^2+v_{\rho^\prime}^2+v_\chi^2+v^2_{\chi^\prime})^2} \right)
 \end{eqnarray}
and  the mass eigenvectors, respectively:
\begin{eqnarray}
A^{\mu}&=&\frac{1}{\sqrt{1+\frac{2t^2}{3}}}\left(\frac{t}{\sqrt{6}}
 V_{3}^{\mu}-\frac{t}{\sqrt{2}}V_{8}^{\mu}+B^{\mu}
 \right),
 \crn
Z^{\mu} &=&
 -\frac{\sqrt{3}(c_\varsigma
+\frac{s_\varsigma}{\sqrt{3+2t^2}})}{2}V_{3}^{\mu}
 -\frac{-c_\varsigma
+\frac{3s_\varsigma}{\sqrt{3+2t^2}}}{2}V_{8}^{\mu}
+\frac{\sqrt{2}ts_{\varsigma}}{\sqrt{3+2t^2}}B^{\mu},
 \crn
Z^{\prime \mu} &=&    -\frac{\sqrt{3}(-s_\varsigma
+\frac{c_\varsigma}{\sqrt{3+2t^2}})}{2}V_{3}^{\mu}
 -\frac{s_\varsigma
+\frac{3c_\varsigma}{\sqrt{3+2t^2}}}{2}V_{8}^{\mu}
-\frac{\sqrt{2}tc_{\varsigma}}{\sqrt{3+2t^2}}B^{\mu},\nn
\end{eqnarray}

 with  $t$ and $\varsigma$  are defined as follows
 \begin{eqnarray}
t &=& \frac{g^\prime}{g} \equiv \frac{6 \sin^{2}\theta_{W}}{1-4 \sin^{2}\theta_{W}},
\label{tisot}\\
\tan ( 2\varsigma ) &=&  \frac{\sqrt{3+2t^2}}{1+t^2} \left(
\frac{v_\chi^2+v_{\chi^\prime}^2
-v_\rho^2-v_{\rho^\prime}^2}{v_\chi^2+v_{\chi^\prime}^2+v_\rho^2+v_{\rho^\prime}^2}
\right).
\end{eqnarray}
The relation in (\ref{tisot})
 predicts that there exists an
energy scale at which the model loses its perturbative character
as we have noted at the main aspect of the 3-3-1 models.
Therefore, in order to  keep  its  perturbative character, we have
 $\sin^{2}\theta_{W}( \mu)<1/4$ at any
energy scale.

Let us summary the gauge mass spectrum. The gauge boson mixing is
separated into two parts. One is charged gauge bosons and one is
neutral gauge bosons. The  exact eigenvectors and eigenvalues are
obtained.
 According to  the limit given in (\ref{cond}), we get the constraint on the gauge mass as follows
\begin{equation}
M_{Z^{\prime}}>M_{U}>M_{V}>M_{Z}>M_{W}.
\end{equation}
This constraint is similar to  those in \cite{massspectrum}. As
all new gauge masses are proportional to $v_{\chi}$ and
$v_{\chi^{\prime}}$, both are in the TeV scale [see
Eq.(\ref{breaksusy331tou1})].  It explains  why the new gauge
bosons have not been yet detected,  but their masses can be
discovered by the experiments at the Large Hadron Collider (LHC)
and at the International Linear Collider (ILC).

\section{Fermion mass matrices}
\label{sec3}

In this section we will show that all the fermions of this model get masses in
concordance with the experimental data.

\subsection{Doubly charged charginos }
 As in previous works \cite{mcr,massspectrum}, we get the same result
without any modification. These new states can be discovered in
the LHC throughout the following \be \bar{p}+p \longrightarrow
\tilde{\chi}^{++}\tilde{\chi}^{--}. \ee The similar one
\cite{massspectrum} \be e^{-}+e^{-} \longrightarrow
\tilde{\chi}^{--}\tilde{\chi}^{0}, \ee is a prospective reaction
in the ILC.

\subsection{Charged leptons and charginos }

Let us consider the mass spectrum of the charged leptons and
charginos.  In the  model under consideration,  mass  mixing
matrix  of the charged leptons and charginos is similar to that
given in \cite{Dong:2006vk,Dong:2007qc,331susy1}.
 In our case, the
$e,\mu$ and $\tau$ leptons gain mass without
 a sextet Higgs or
the charged lepton singlet. Note that the Higgsinos
$\tilde{\rho},\tilde{\chi}$ and their respective primed fields
have the same charge assignment of the triplets $\rho$ and $\chi$.
Hence, they can mix  with the usual leptons.

Let us first consider  the charged lepton and chargino masses.
Denoting
\begin{equation}
\begin{array}{c}
\phi^+=(e^c,\mu^c,\tau^c,-i\lambda^+_W,-i\lambda^+_V,
\tilde{\rho}^+,\tilde{\chi}^{\prime+},\tilde{\rho^\prime}^+, \tilde{\chi}^+)^T,\\
\phi^-=(e,\mu,\tau,-i\lambda^-_W,-i\lambda^-_V,
\tilde {\rho}^{\prime -},\tilde{\chi}^{-},\tilde{\rho^\prime}^-, \tilde{\chi}^-)^T,
\end{array}
\label{cbasis}
\end{equation}
where all the fermionic fields are still Weyl spinors, we can
also, as before, define $\Psi^{\pm}=(\phi^+ \phi^-)^T$.  Then, the
mass term  is written in the form $-(1/2)[\Psi^{\pm T}$
$Y^\pm\Psi^{\pm}+\mathrm{H.c.}]$ where $Y^\pm$ is given by:
\begin{equation}
Y^{\pm}= \left( \begin{array}{cc}
0  & X^T \\
X  & 0
\end{array}
\right),
\label{ypm}
\end{equation}
with $X$ matrix defined as
\begin{eqnarray}
X= \left( \begin{array}{ccccccccc}
0& 0& 0& 0& 0&  \frac{\lambda_{2e}}{3}w& 0&0&0\\
0& 0& 0& 0& 0&  \frac{\lambda_{2 \mu}}{3}w& 0&0&0\\
0& 0& 0& 0& 0&  \frac{\lambda_{2 \tau}}{3}w& 0&0&0\\
0& 0& 0&m_\lambda& 0& gu& 0& -gu^\prime&0\\
0& 0& 0& 0&m_{\lambda}&0& -gw^\prime & 0&gw\\
 \frac{\lambda_{2e}}{3}w& \frac{\lambda_{2 \mu}}{3}w&  \frac{\lambda_{2 \tau}}{3}w&
gu&
  0&0& 0&- \frac{\mu_{\rho}}{2}&0\\
0&0& 0& 0&- gw^\prime&0&0& 0&-\frac{ \mu_\chi}{2}\\
0&0&0&-g u^\prime&0&-\frac{\mu_\rho}{2}&0&0&0 \\
0&0&0&0 &gw&0&-\frac{ \mu_\chi}{2}&0&0
\end{array}
\right),
\end{eqnarray}
This mass matrix gives two  zero   eigenvalues \cite{331susy1}.
One of two zero eigenvalues is identified to the electron mass and
the remaining one is identified to the muon mass. It means that
the electron and muon are massless at the tree level. If there is
not a discrete symmetry, which is  added to  the Lagrangian, the
charged lepton can get a mass by loop corrections as done in
\cite{cmmc}.

\begin{figure}[h]
\begin{center}
\vglue -0.009cm
\centerline{\epsfig{file=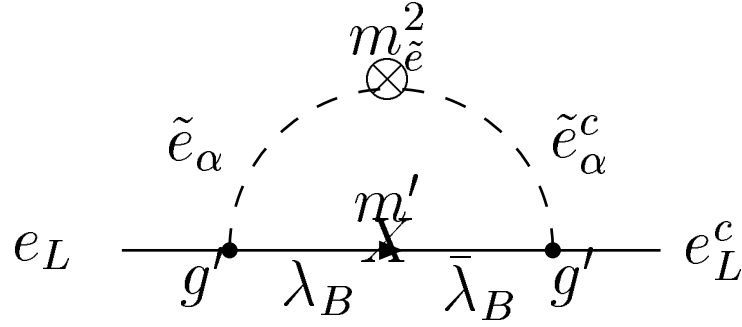,width=0.7\textwidth,angle=0}}
\end{center}
\caption{Diagram  giving  mass to electron which does not appear
in the superpotential. The diagram to the muon is similar (just change the
selecton by the smuon)} \label{fig1}
\end{figure}

In this model, the electron still couples with the gaugino
$\lambda_{B}$ of $\mathrm{U(1)}$ group, [see Eq.(\ref{llchar})],
in a similar way as shown in \cite{cmmc}.  As the selectrons and
the gauginos get their masses due the soft terms given
 in  Eqs.(\ref{gmt1},\ref{smt}), it allows us to draw
the diagram of Fig.\ref{fig1} that gives
contribution to the electron mass. Therefore, at the one loop correction the electron mass is given by:
\begin{eqnarray}
m_{e}  & \propto&
\frac{\alpha_{\mathrm{U(1)}}\sin(2\theta_{\tilde{e}})}{\pi}
m^{\prime}\left[
\frac{m_{\tilde{e_{1}}}^{2}}{m_{\tilde{e_{1}}}^{2}-m^{\prime2}}
\ln\left(  \frac{m_{\tilde{e_{1}}}^{2}}{m^{\prime2}}\right)
\right. \crn & -& \left.
\frac{m_{\tilde{e_{2}}}^{2}}{m_{\tilde{e_{2}}}^{2}-m^{\prime2}}
\ln\left(  \frac{m_{\tilde{e_{2}}}^{2}}{m^{\prime2}}\right)
\right]  \,\ , \crn m_{\mu}  & \propto&
\frac{\alpha_{\mathrm{U(1)}}\sin(2\theta_{\tilde{\mu}})}{\pi}
m^{\prime}\left[
\frac{m_{\tilde{\mu_{1}}}^{2}}{m_{\tilde{\mu_{1}}}^{2}-m^{\prime2}}
\ln\left(  \frac{m_{\tilde{\mu_{1}}}^{2}}{m^{\prime2}}\right)
\right. \crn & -& \left.
\frac{m_{\tilde{\mu_{2}}}^{2}}{m_{\tilde{\mu_{2}}}^{2}-m^{\prime2}}
\ln\left(  \frac{m_{\tilde{\mu_{2}}}^{2}}{m^{\prime2}}\right)
\right]  \,\ , \label{emasscorr}
\end{eqnarray}
where $m^\prime, m_{\tilde{e}}, m_{\tilde{\mu}}$ are soft
parameters given in Eq.~(\ref{gmt1}),
$\alpha_{\mathrm{U(1)}}=g^{\prime2}/(4\pi)$ and the
$\theta_{\tilde{e}}, \theta_{\tilde{\mu}}$   are  defined in
Eq.(\ref{eq:Rsq}). It is the simplest way to generate fermion
masses through one-loop correction \cite{cmmc,banks,ma}. As we
expect the smuon is heavier than the selectron, it explains why
the muon is heavier than electron, see at SPS scenarios
\cite{sps1,sps2,sps}.  However, ones can worry about the following
fact: the supersymmetric masses are strongly constrained by recent
LHC data \cite{atlasgluinos}. Thus, it is not clear at all that
the  obtained fermion masses are correct. Fortunately in this
model there appear two sources that the lepton particles obtained
mass  through radiative
 mechanism.
 We will below comment in brief these
mechanisms.

\begin{figure}[h]
\begin{center}
\includegraphics[width=8cm]{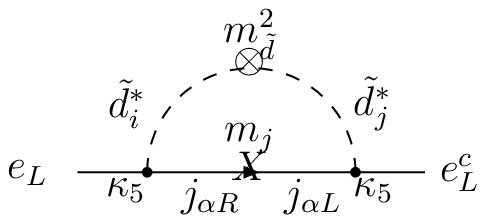}
\end{center}
\caption{Diagram generating the electron mass taking into account the coupling $\kappa_{5}$. }
\label{figeletrona}
\end{figure}

\begin{figure}[h]
\begin{center}
\includegraphics[width=8cm]{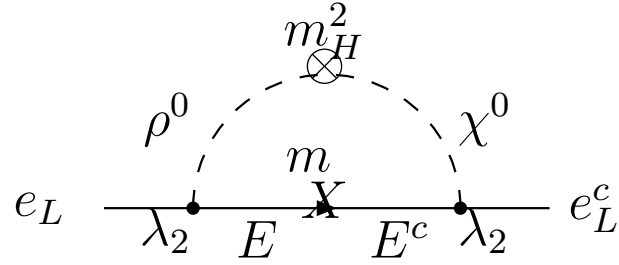}
\end{center}
\caption{Diagram generating the electron mass taking into account
the coupling $\lambda_{2}$ with $E, E^c$ are the singlet charged
Higgsinos.} \label{figeletronb}
\end{figure}

One of these sources is the coupling $\kappa_{5\alpha ij}$, from
Eq.(\ref{sp1}), that generate the diagram of Fig.\ref{figeletrona}
(for more details,
 see \cite{331susy1}).
The second mechanism is  the coupling $\lambda_{2}$, from Eq.(\ref{sp1}).
 This contribution will
generate four diagrams, we draw only one of these possibilities in
Fig.\ref{figeletronb}. We would like  to stress that the diagrams
given in Fig.\ref{figeletrona} and Fig.\ref{figeletronb} are the
sources of the non-renormalizable interaction $(L \chi)(L \rho)$
that appears in the non-supersymmetric model. Of course, from
these contributions we can get the new expression to the electron
(and also to the muon) masses.
 They are very similar to ones given at Eq.(\ref{emasscorr}), but more larger. We
will not write them here. We can now show, including all the contributions the electron and the muon
will get masses.

The tau gets its mass at tree level, it explains why the tau is
heavier than the muon and electron. The other four mass values are
at $\rm GeV$ scale as shown in \cite{331susy1}. The way we perform
the diagonalization, as well, the particle definitions are given
in \cite{Dong:2006vk,Dong:2007qc,331susy1}.

\subsection{Neutrinos and neutralinos}

\begin{figure}[ht]
\begin{center}
\vglue -0.009cm
\centerline{\epsfig{file=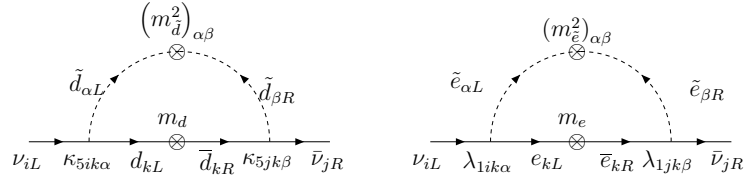,width=0.7\textwidth,angle=0}}
\end{center}
\caption{ Diagrams  giving masses to electron's and muon's
neutrinos which do not appear in the superpotential, $\tilde{e}$
is the selectron and $\tilde{d}$ is the down-squark and the label
$\alpha =1,2$.} \label{fig2}
\end{figure}

Because the existence of the interaction between neutrinos and neutralinos, their mass
matrix has a mixture. The mass term in the basis

\begin{equation}
\Psi^{0}=\left(%
\begin{array}{cccccccccc}
  \nu_{e} &\nu_{\mu} & \nu_{\tau}
   & -i \lambda^{3}_{A} & -i \lambda^{8}_{A} & -i \lambda_{B}  & \tilde{\rho}^{0}
   & \tilde{\rho}^{ \prime 0}
   &\tilde{\chi}^{0} &
   \tilde{\chi}^{ \prime 0}\\
\end{array}%
\right)^T
\end{equation}
is given by $-(1/2)[ \left( \Psi^{0} \right)^T Y^{0}
\Psi^{0}+\mathrm{H.c.}] $,  where
\begin{equation}
\footnotesize{Y^{0}= \left( \begin{array}{cccccccccc}
0& 0& 0& 0& 0& 0& \frac{ \lambda_{2e}}{3}w& 0& \frac{ \lambda_{2e}}{3}u& 0\\
0& 0& 0& 0& 0& 0& \frac{ \lambda_{2 \mu}}{3}w& 0&
\frac{ \lambda_{2\mu}}{3}u& 0\\
0& 0& 0& 0& 0& 0& \frac{ \lambda_{2 \tau}}{3}w& 0&
\frac{ \lambda_{2 \tau}}{3}u& 0\\
0& 0& 0& m_{\lambda}& 0& 0& -
\frac{gu}{\sqrt{2}}& \frac{gu^{\prime}}{\sqrt{6}}& 0& 0\\
0& 0& 0& 0& m_{\lambda}& 0& \frac{gu}{\sqrt{6}}&- \frac{gu^{\prime}}{\sqrt{6}}&\frac{gw}{\sqrt{6}} &
-\frac{gw^\prime}{\sqrt{6}}\\
0& 0& 0& 0& 0& m^{\prime}& \frac{g^{\prime}u}{\sqrt{6}}&-
\frac{g^{\prime}u^{\prime}}{\sqrt{6}}& -\frac{g^{\prime}w}{\sqrt{6}}&
\frac{g^{\prime}w^{\prime}}{\sqrt{6}}\\
\frac{ \lambda_{2e}}{3}w& \frac{ \lambda_{2 \mu}}{3}w&
\frac{ \lambda_{2 \tau}}{3}w&-
\frac{gu}{\sqrt{2}}& \frac{gu}{\sqrt{6}}& \frac{g^{\prime}u}{\sqrt{2}}& 0&-
\frac{ \mu_{\rho}}{2}& 0& 0\\
0& 0& 0&  \frac{gu^{\prime}}{\sqrt{2}}&- \frac{gu^{\prime}}{\sqrt{6}}&
-\frac{g^{\prime}u^{\prime}}{\sqrt{6}}&- \frac{ \mu_{\rho}}{2}& 0& 0& 0\\
\frac{ \lambda_{2e}}{3}u& \frac{ \lambda_{2\mu}}{3}u&
\frac{ \lambda_{2 \tau}}{3}u& 0&
 \frac{gw}{\sqrt{6}} &\frac{g^{\prime}w}{\sqrt{6}}& 0& 0& 0&- \frac{ \mu_{\chi}}{2}\\
0& 0& 0& 0&- \frac{ gw^{\prime}}{\sqrt{6}}& \frac{g^{\prime}w^{\prime}}{\sqrt{6}}& 0& 0&-
\frac{ \mu_{\chi}}{2}& 0
\end{array}
\right).} \label{mmn}
\end{equation}
This matrix  has  two zero eigenvalues. The
 parameter $m_\lambda$ is defined in
Eq.~(\ref{gmt1}).  On the other hand, the electron's neutrino
still couples with the selectron and down-squark, while the muon's
neutrinos couple with smuons and strange squark. These couplings
lead to the diagrams shown in  Fig.\ref{fig2}.
These diagrams give the contribution to the neutrino mass
given in Eq.(\ref{eletronneutrino}).
\begin{scriptsize}
{\footnotesize •}
\end{scriptsize} \begin{eqnarray}
m_{\nu_{e}} &\simeq&  \frac{1}{8 \pi^{2}}\sum_{\imath =1}^{2} \left \{ m^{\prime}
\left[  \lambda_{1e11}\lambda_{1e11}\frac{m_{\tilde{e_{1}}}^{2}}{m_{\tilde{e_{1}}}^{2}-m^{\prime2}}
\ln\left(  \frac{m_{\tilde{e_{1}}}^{2}}{m^{\prime2}}\right)  \right. \right.
\crn
&-&  \left. \lambda_{1e12}\lambda_{1e12}  \frac{m_{\tilde{e_{2}}}^{2}}{m_{\tilde{e_{2}}}^{2}-m^{\prime2}}
\ln\left(  \frac{m_{\tilde{e_{2}}}^{2}}{m^{\prime2}}\right)  \right]
\crn &+& 3
m_{\tilde{g}}
\left[ \kappa_{5e11}\kappa_{5e11}  \frac{m_{\tilde{d_{1}}}^{2}}{m_{\tilde{d_{1}}}^{2}-m^{\prime2}}
\ln\left(  \frac{m_{\tilde{d_{1}}}^{2}}{m^{2}_{\tilde{g}}}\right)  \right.
\crn
&-& \left. \left. \kappa_{5e12}\kappa_{5e12}  \frac{m_{\tilde{d_{2}}}^{2}}{m_{\tilde{d_{2}}}^{2}-m^{\prime2}}
\ln\left(  \frac{m_{\tilde{d_{2}}}^{2}}{m^{2}_{\tilde{g}}}\right)  \right]
\right \} , \crn
m_{\nu_{\mu}} &\simeq&  \frac{1}{8 \pi^{2}}\sum_{\imath =1}^{2} \left \{ m^{\prime}
\left[  \lambda_{1\mu 21}\lambda_{1\mu 21}\frac{m_{\tilde{\mu_{1}}}^{2}}{m_{\tilde{\mu_{1}}}^{2}-m^{\prime2}}
\ln\left(  \frac{m_{\tilde{\mu_{1}}}^{2}}{m^{\prime2}}\right)  \right. \right.
\crn
&-&  \left. \lambda_{1\mu 22}\lambda_{1\mu 22}  \frac{m_{\tilde{\mu_{2}}}^{2}}{m_{\tilde{\mu_{2}}}^{2}-m^{\prime2}}
\ln\left(  \frac{m_{\tilde{\mu_{2}}}^{2}}{m^{\prime2}}\right)  \right]
\crn &+& 3
m_{\tilde{g}}
\left[ \kappa_{5\mu 21}\kappa_{5\mu 21}  \frac{m_{\tilde{s_{1}}}^{2}}{m_{\tilde{s_{1}}}^{2}-m^{\prime2}}
\ln\left(  \frac{m_{\tilde{s_{1}}}^{2}}{m^{2}_{\tilde{g}}}\right)  \right.
\crn
&-& \left. \left. \kappa_{5\mu 22}\kappa_{5\mu 22}  \frac{m_{\tilde{s_{2}}}^{2}}{m_{\tilde{s_{2}}}^{2}-m^{\prime2}}
\ln\left(  \frac{m_{\tilde{s_{2}}}^{2}}{m^{2}_{\tilde{g}}}\right)  \right]
\right \} ,
\label{eletronneutrino}
\end{eqnarray}
where $m_{\tilde{g}}$ is the mass of a gluino,
$m_{\tilde{d}}$ is the down-squark mass
  and $m_{\tilde{s}}$ is the strange-squark mass.

The neutrino masses are proportional to $\lambda_1, \kappa_5$, the
parameters which  break the lepton number conservation. The
electron mass is proportional to $g$. Due to this fact we expect
that it must be satisfied a condition $\lambda_1, \kappa_5 \ll
g^\prime$, then we can explain the reason why neutrinos are much
lighter than the charged leptons.

\begin{figure}[h]
\begin{center}
\includegraphics[width=8cm]{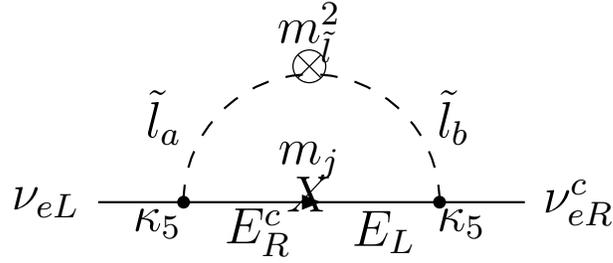}
\end{center}
\caption{Diagram generating the electron's neutrino mass taking into account the coupling $\kappa_{5}$. }
\label{figneutrinoa}
\end{figure}

\begin{figure}[h]
\begin{center}
\includegraphics[width=8cm]{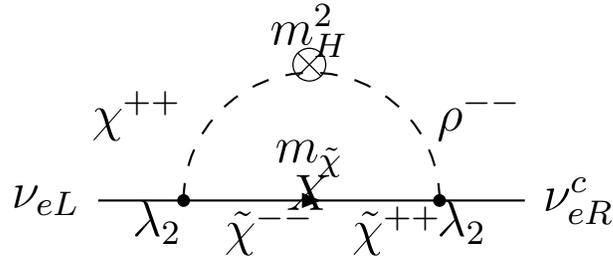}
\end{center}
\caption{Diagram generating the electron's neutrino mass taking into account the coupling $\lambda_{2}$.}
\label{figneutrinob}
\end{figure}

 As happened with the charged leptons, we can generate new contribution
to neutrino masses in an analogous way as done to the charged
leptons. We draw the new contributions in Fig.\ref{figneutrinoa}
and Fig.\ref{figneutrinob}. Therefore, as in the previous case,
including all the contributions, the electron's and muon's
neutrinos will get their masses can be satisfied the experimental
data.

\subsection{Quarks}

\begin{figure}[h]
\begin{center}
\vglue -0.009cm
\centerline{\epsfig{file=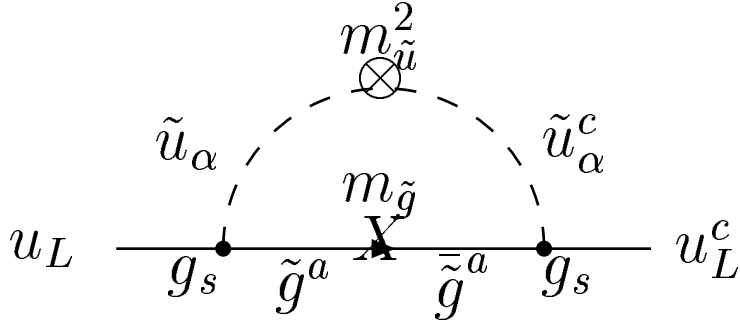,width=0.7\textwidth,angle=0}}
\end{center}
\caption{ Diagram  giving mass to up quark
 which does not appear in the superpotential, $\tilde{u}$ is the up-squark.}
\label{fig3}
\end{figure}

Let us first consider the u-quark type.  First, we define the basis as done in
Ref. \cite{massspectrum}, particularly
\begin{equation}
\begin{array}{c}
\psi _{u}^{+}= \left( \begin{array}{ccc} u_{1}&
                                         u_{2}&
                                         u_{3}
\end{array} \right)^{T}, \,\
\psi _{u}^{-}= \left( \begin{array}{ccc} u^{c}_{1}&
                                         u^{c}_{2}&
                                         u^{c}_{3}
\end{array} \right)^{T},
\end{array}
\label{cbasis}
\end{equation}
where all the u-quark fields are still Weyl spinors, we
can also define \newline $\Psi _{u}^{\pm }$ = $(\psi _{u}^{+} \psi
_{u}^{-}) ^{T}$.   Then,   the mass term is written   in the
form  $$-(1/2)[ \Psi _{u}^{\pm T}Y_{u}^{ \pm}\Psi _{u}^{\pm } +
\mathrm{H.c.}].$$  Here $Y_{u}^{\pm}$ is given by
\begin{equation}
Y_{u}^{ \pm}=
\left(
\begin{array}{cc}
0 & X_{u}^{T} \\
X_{u} & 0
\end{array}
\right),
\label{ypm}
\end{equation}
with
\begin{equation}
X_{u}=\frac{1}{3}\left(
\begin{array}{ccc}
\kappa_{311}u &- \kappa_{312}u & 0 \\
-\kappa_{321}u & \kappa_{322}u & 0 \\
-\kappa_{331}u & \kappa_{323}u & 0 \\
\end{array}
\right),
\label{clmm}
\end{equation}
where the VEVs are defined in Eq.(\ref{vev1}). The mass spectrum
of the up quarks contains one massless particle. However the
lightest quark will get mass due its coupling to gluino as shown
in  Fig.\ref{fig3}. Therefore the up quark's mass is given by
\begin{eqnarray}
m_{u}  \propto
\frac{\alpha_{s}\sin(2\theta_{\tilde{u}})}{\pi}m_{\tilde{g}}
\left[
\frac{M_{\tilde{u_{1}}}^{2}}{M_{\tilde{u_{1}}}^{2}-m_{\tilde{g}}^{2}}
\ln\left(  \frac{M_{\tilde{u_{1}}}^{2}}{m_{\tilde{g}}^{2}}\right)
-
\frac{M_{\tilde{u_{2}}}^{2}}{M_{\tilde{u_{2}}}^{2}-m_{\tilde{g}}^{2}}
\ln\left(  \frac{M_{\tilde{u_{2}}}^{2}}{m_{\tilde{g}}^{2}}\right)
\right]  \,\ , \label{umass}
\end{eqnarray}
where $\al_{s}=g^2_{s}/(4\pi)$, $m_{\tilde{g}}$ and
$m_{\tilde{u}}$ are  the masses of the gluino and
up-squark, respectively, and $\theta_{\tilde{u}}$ is the mixing
angle of left- and right-handed up-squarks given in
Eq.(\ref{eq:Rsq}).

Let us consider the d-quark type.   Doing similarly  as in the
up-quark sector,  we  define $\Psi _{d}^{\pm }=(\psi _{d}^{+} \psi
_{d}^{-})^{T}$,  then  the mass term is written in the form
$-(1/2)[ \Psi _{d}^{\pm T}Y_{d}^{ \pm}\Psi _{d}^{\pm
}+\mathrm{H.c.}]$ where $Y_{d}^{ \pm}$ is given by \be Y_{d}^{
\pm}= \left(
\begin{array}{cc}
0 & X_{d}^{T} \\
X_{d} & 0
\end{array}
\right),
\label{ypmd}
\ee
with
\be
X_{d}=\frac{1}{3}\left(
\begin{array}{ccc}
0 & 0 & 0 \\
0 & 0 & 0 \\
\kappa_{11}u^{\prime} & \kappa_{12}u^{\prime} & \kappa_{13}u^{\prime}
\end{array}
\right), \label{cdmm} \ee In this sector,  there are two massless
eigenvalues.
 We can implement the
same mechanism analyzed in \cite{cmmc1} to give mass to $d$ and
$s$ quarks. Thus,  the model under consideration  is compatible
with chiral theory.

\begin{figure}[h]
\begin{center}
\vglue -0.009cm
\centerline{\epsfig{file=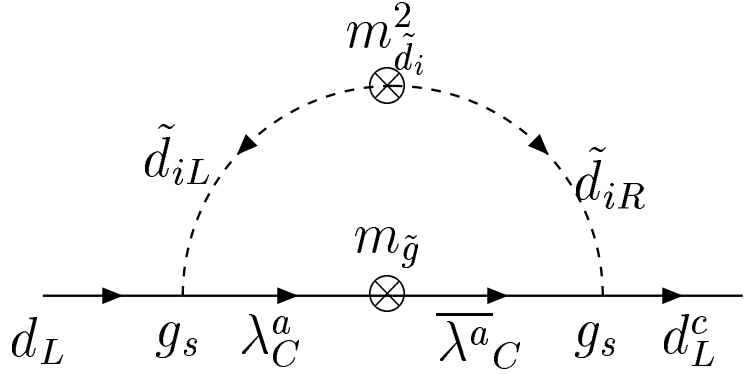,width=0.7\textwidth,angle=0}}
\end{center}
\caption{ Diagram  giving mass to quark $d$ which does not appear
in the superpotential, $\tilde{g}$ is the gluino, $\tilde
{d}_{i}$, $i=1,2$, is the down-squark.} \label{fig4}
\end{figure}

\begin{figure}[h]
\begin{center}
\vglue -0.009cm
\centerline{\epsfig{file=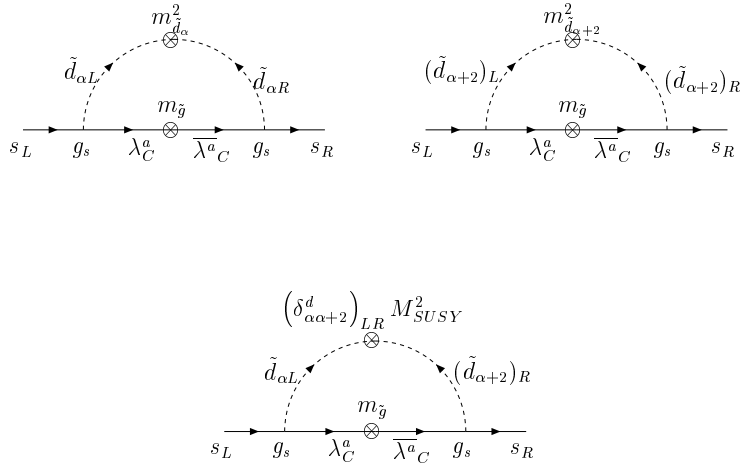,width=0.7\textwidth,angle=0}}
\end{center}
\caption{ Diagram  giving  mass to $s$ quark  which does not
appear in the superpotential, $\tilde{g}$ is the gluino,
$\tilde {s}_{i}$ and $\tilde{b}_{i}$, $i=1,2$, are the squark $s$
and sbottom,
respectively.}
\label{fig5}
\end{figure}

Analogously,   looking at
Fig.\ref{fig4}  and Eq.(\ref{eq:Rsq}), we get the expression for   mass of $d$-quark
\cite{cmmc}
\bea
m_{d} \propto \frac{\alpha_{s}\sin(2\theta_{\tilde{d}})}{\pi}
m_{\tilde{g}}\left[
\frac{M_{\tilde{d_{1}}}^{2}}{M_{\tilde{d_{1}}}^{2}
-m_{\tilde{g}}^{2}}\ln\left(
\frac{M_{\tilde{d_{1}}}^{2}}{m_{\tilde{g}}^{2}}\right) -
\frac{M_{\tilde{d_{2}}}^{2}}{M_{\tilde{d_{2}}}^{2}-
m_{\tilde{g}}^{2}}\ln\left(
\frac{M_{\tilde{d_{2}}}^{2}}{m_{\tilde{g}}^{2}}\right) \right] .
\label{dmass1}
\eea
For the  $s$-quark, looking at  Fig.\ref{fig5} and Eq.(\ref{squarkbs}), we obtain \cite{cmmc1}
\bea
m_{s} &=&\frac{\al_{s} m_{\tilde{g}}}{4\pi ^{3}}\sum_{\alpha
=1}^{2} \left\{ R_{1\alpha}^{(d)}R_{2\alpha }^{(d)}
\frac{m_{\tilde{g}}^{2}}{(m_{\tilde{g}}^{2}-m_{\tilde{d}_{\alpha
}}^{2})} \ln \left( \frac{m_{\tilde{g}}^{2}}{m_{\tilde{d}_{\alpha
}}^{2}}\right) \right.\crn&+&\left. R_{1\alpha
+2}^{(d)}R_{2\alpha +2}^{(d)}
\frac{m_{\tilde{g}}^{2}}{(m_{\tilde{g}}^{2}-m_{\tilde{d}_{\alpha
+2}}^{2})} \ln \left(
\frac{m_{\tilde{g}}^{2}}{m_{\tilde{d}_{\alpha +2}}^{2}}\right)
\right.   \crn
&+& \left.
\frac{R_{1\alpha }^{(d)}R_{2\alpha +2}^{(d)}}
{(m_{\tilde{d}_{\alpha }}^{2}-m_{\tilde{d}_{\alpha +2}}^{2})
(m_{\tilde{g}}^{2}-m_{\tilde{d}_{\alpha }}^{2})(m_{\tilde{d}_{\alpha +2}}^{2}-m_{\tilde{g}}^{2})}
\right.   \crn
&\times& \left. \left( \delta_{12}^{d}\right)_{LR}M_{SUSY}^{2}
\left[ m_{\tilde{d}_{\alpha }}^{2}m_{\tilde{d}_{\alpha +2}}^{2}
\ln \left( \frac{m_{\tilde{d}_{\alpha }}^{2}}{m_{\tilde{d}_{\alpha
+2}}^{2}}
\right) \right. \right. \crn
&+& \left. \left. m_{\tilde{d}_{\alpha }}^{2}m_{\tilde{g}}^{2} \ln
\left( \frac{m_{\tilde{g}}^{2}}{m_{\tilde{d}_{\alpha }}^{2}}
\right) + m_{\tilde{d}_{\alpha +2}}^{2}m_{\tilde{g}}^{2} \ln
\left( \frac{m_{\tilde{d}_{\alpha +2}}^{2}}{m_{\tilde{g}}^{2}}
\right) \right] \right\},\label{smass1}
\eea
 where $\theta_{\tilde{d}}$ are mixing angles, $R_{\beta \alpha }^{d}$ is
defined at Eq.(\ref{squarkbs}) and
$m_{\tilde{d}_{\alpha }}^{2}$ are the eigenvalues of
Eq.(\ref{eq.usquarkmass}) and they  are the physical masses of $\tilde
{s}_{1},\tilde {s}_{2}$, $\tilde{b}_{1}$ and $\tilde{b}_{2}$.

The electron mass is given by Eq.(\ref{emasscorr}), the mass of up
quarks is given by Eq.(\ref{umass}) and the down quark
is given by Eq.(\ref{dmass1}). Note that the quark masses
are proportional to $g_{s}$ while lepton masses are proportional
to $g$.  The fact that $g_{s}\gg g$ gives an explanation why
quarks are heavier than the leptons. The mass of $s$-quark is
given by Eq.(\ref{smass1}), comparing this formula with
Eq.(\ref{dmass1}) we can explain why $s$-quark is heavier
than $d$-quark.

\subsection{Sfermions}

It is  known that in the general case, the sfermions  have a
 flavor mixing. It leads to  all sfermions mass matrices  are
$6 \times 6$ matrices \cite{cmmc1}. Therefore the slepton
sector contains lepton flavor violation at the tree level. In
order to avoid this problem we will neglect the generation mixing
in  the slepton sector.  This assumption is not held for all other
squark sectors.
 Each $6 \times 6$ slepton mass matrix  can be divided into
 three $2 \times 2$ mass matrices.
The off diagonal left-right mixing is proportional to the fermion masses.

Here, we will only  present the main formulas. In the case of
charged sleptons we can  generally  write $2 \times 2$ mass
matrices \be \mathcal{M}_{\tilde{f}}^{2} = \left(
\begin{array}{cc}
m_{\tilde{f}_{L}}^{2} & a_{f} m_{f} \\
a_{f} m_{f} & m_{\tilde{f}_{R}}^{2}
\end{array}
\right) \;=\; (\mathcal{R}^{\tilde{f}}) \left(
\begin{array}{cc}
m_{\tilde{f}_{1}}^{2} & 0 \\
0 & m_{\tilde{f}_{2}}^{2}
\end{array}
\right) \mathcal{R}^{\tilde{f}},  \label{eq:msqmat} \ee where
$\tilde{f}= \tilde{e},~ \tilde{\mu},~ \tilde{\tau}$. The weak
eigenstates $\tilde{f}_{L}$ and $\tilde{f}_{R}$ are thus related
to their mass eigenstates $\tilde{f}_{1} $ and $\tilde{f}_{2}$,
where $\tilde{f}_{1}$ is the lighter sfermion, by \be \left(
\begin{array}{c}
\tilde{f}_{1} \\
\tilde{f}_{2}
\end{array}
\right) = \mathcal{R}^{\tilde{f}} \, \left(
\begin{array}{c}
\tilde{f}_{L} \\
\tilde{f}_{R}
\end{array}
\right), \hspace{8mm} \mathcal{R}^{\tilde{f}} = \left(
\begin{array}{rr}
\cos\theta_{\tilde{f}} & \sin\theta_{\tilde{f}} \\
-\sin\theta_{\tilde{f}} & \cos\theta_{\tilde{f}}
\end{array}
\right) ,  \label{eq:Rsq}
\ee
with $\theta_{\tilde{f}}$ is  the slepton mixing angle.

The four component vectors for up-squarks and down-squarks are,
respectively, ($\tilde{u}_{1 L}$, $\tilde{u}_{2L}$,
$\tilde{u}_{1R}$, $\tilde{u}_{2R}$) and ($\tilde{d}_{1L}$,
$\tilde{d}_{2L}$, $\tilde{d}_{1R}$, $\tilde{d}_{2R}$). Thus, the
squark squared mass matrices are given by
\begin{equation}
\mathcal{M}_{\tilde{u}, \tilde{d}}^{2}=\left(
\begin{array}{llll}
M_{\tilde{L},c\{s\}}^{2} & (M_{\tilde{U}\{\tilde{D}\}}^{2})_{LL} &
m_{c\{s\}}
\mathcal{A}_{c\{s\}} & (M_{\tilde{U}\{\tilde{D}\}}^{2})_{LR} \\
(M_{\tilde{U}\{\tilde{D}\}}^{2})_{LL} & M_{{\tilde{L}}t\{b\}}^{2} &
(M_{\tilde{U}\{\tilde{D}\}}^{2})_{RL} &
m_{t\{b\}}\mathcal{A}_{t\{b\}} \\
(M_{\tilde{U}\{\tilde{D}\}}^{2})_{LR} & (M_{\tilde{U}\{\tilde{D}\}}^{2})_{RL}
& M_{{\tilde{R}}c\{s\}}^{2} & (M_{\tilde{U}\{\tilde{D}\}}^{2})_{RR} \\
(M_{\tilde{U}\{\tilde{D}\}}^{2})_{LR} &
m_{t\{b\}}\mathcal{A}_{t\{b\}} &
(M_{\tilde{U}\{\tilde{D}\}}^{2})_{RR} & M_{{\tilde{R}}t\{b\}}^{2}
\end{array}
\right) .  \label{eq.usquarkmass}
\end{equation}

In order to diagonalize
$\mathcal{M}_{\tilde{u}\{\tilde{d}\}}^{2}$, two rotation $4\times
4$ matrices, $R^{(u)}$ and $R^{(d)}$, one for the $up$-squarks and
the other for $down$-squarks, are needed. Thus the squark mass
eigenstates ($\tilde{q}_{\alpha }^{\prime }$) and the  weak
 squark eigenstates ($\tilde{q}_{\alpha }$) are related by
\begin{equation}
\tilde{q}_{\alpha }^{\prime }=\sum R_{\alpha \beta }^{(q)}
\tilde{q}_{\beta}\,.
\label{squarkbs}
\end{equation}
One obtains the squark mass eigenvalues and eigenstates after the
diagonalization procedure as indicated in Ref.~\cite{herrero}.

\subsection{Gluinos, exotic quarks and sfermions}

For the exotic quarks and gluinos, their masses are the same
 as presented at \cite{massspectrum}.

\section{Higgs potential}
\label{sec:sp}

As usual, the scalar Higgs potential is written as
\be
V_{3-3-1}=V_{\mathrm{D}}+V_{\mathrm{F}}+V_{\mathrm{soft}}
\label{ep1}
\ee
 with
\bea
V_{\mathrm{D}}&=&-\mathcal{L}_{D}=\frac{1}{2}\left(D^{a}D^{a}+DD\right)\crn
&=& \frac{g^{\prime
2}}{12}(\bar{\rho}\rho-\bar{\rho^{\prime}}\rho^{\prime}
-\bar{\chi}\chi+\bar{\chi^{\prime}}\chi^{\prime})^{2} \crn &+&
\frac{g^{2}}{8}\sum_{i,j}\left(\bar{\rho}_{i}\lambda^{a}_{ij}\rho_{j}
+\bar{\chi}_{i}\lambda^{a}_{ij}\chi_{j}-
\bar{\rho^{\prime}}_{i}\lambda^{*a}_{ij}\rho^{\prime}_{j}
-\bar{\chi^{\prime}}_{i}\lambda^{*a}_{ij}\chi^{\prime}_{j}
\right)^{2}, \crn
V_{\mathrm{F}}&=&-\mathcal{L}_{F}=\sum_{F}\bar{F}_{\mu}F_{\mu}\crn
\\ &=& \sum_{i}\left[ \left\vert
\frac{\mu_{\rho}}{2}\rho^{\prime}_{i}\right\vert^{2}+ \left\vert
\frac{\mu_{\chi}}{2}\chi^{\prime}_{i}\right\vert^{2} + \left\vert
\frac{\mu_{\rho}}{2}\rho_{i} \right\vert^{2}+ \left\vert
\frac{\mu_{\chi}}{2}\chi_{i} \right\vert^{2} \right], \crn
V_{\mathrm{soft}}&=&-\mathcal{L}_{\mathrm{SMT}}=
m^{2}_{\rho}\bar{\rho}\rho+ m^{2}_{\chi}\bar{\chi}\chi+
m^{2}_{\rho^{\prime}}\bar{\rho^{\prime}}\rho^{\prime}+
m^{2}_{\chi^{\prime}}\bar{\chi^{\prime}}\chi^{\prime}, \label{ess}
\eea where  $m^{2}_{\rho}, m^{2}_{\chi}, m^{2}_{\rho^{\prime}},
m^{2}_{\chi^{\prime}}$
 have the mass dimension.

All the four neutral scalar components
$\rho^0,\chi^0,\rho^{\prime0},\chi^{\prime0}$
gain non-zero vacuum expectation values.  Expansions of
the neutral scalars around their VEVs are usually
\begin{eqnarray}
< \rho >&=&\frac{1}{\sqrt{2}}
      \left( \begin{array}{c} 0 \\
v_{\rho}+H_{\rho}+iF_{\rho} \\
                  0          \end{array} \right), \,
< \rho^{\prime} >=\frac{1}{\sqrt{2}}
      \left( \begin{array}{c} 0 \\
v_{\rho^{\prime}}+H_{\rho^{\prime}}+iF_{\rho^{\prime}}\\
                  0          \end{array} \right), \, \crn
< \chi >&=&\frac{1}{\sqrt{2}}
      \left( \begin{array}{c} 0 \\
                  0 \\
v_{\chi}+H_{\chi}+iF_{\chi} \end{array} \right),\,
< \chi^{\prime} >=\frac{1}{\sqrt{2}}
      \left( \begin{array}{c} 0 \\
                  0 \\
v_{\chi^{\prime}}+H_{\chi^{\prime}}+iF_{\chi^{\prime}}\end{array} \right).
\crn
\label{develop}
\end{eqnarray}
Due to the requirement  that the potential to reach a minimum at
the chosen VEV's, which is equivalent to the condition of absence
of the   linear terms in fields, we get a system of constraint
equations \bea && 12m^{2}_{\rho}+3 \mu^{2}_{\rho}+ g^{2} \left(
2v^{2}_{\rho}-2v^{\prime 2}_{\rho}-v^{2}_{\chi}+v^{\prime
2}_{\chi}\right)+ g^{\prime 2}\left( v^{2}_{\rho}-v^{\prime
2}_{\rho}-v^{2}_{\chi}+v^{\prime 2}_{\chi}\right)=0, \crn &&
12m^{2}_{\rho^{\prime}}+3 \mu^{2}_{\rho}- g^{2} \left(
2v^{2}_{\rho}-2v^{\prime 2}_{\rho}-v^{2}_{\chi}+v^{\prime
2}_{\chi}\right)- g^{\prime 2}\left( v^{2}_{\rho}-v^{\prime
2}_{\rho}-v^{2}_{\chi}+v^{\prime 2}_{\chi}\right)=0, \crn &&
12m^{2}_{\chi}+3 \mu^{2}_{\chi}- g^{2} \left(
v^{2}_{\rho}-v^{\prime 2}_{\rho}-2v^{2}_{\chi}+2v^{\prime
2}_{\chi}\right)- g^{\prime 2}\left( v^{2}_{\rho}-v^{\prime
2}_{\rho}-v^{2}_{\chi}+v^{\prime 2}_{\chi}\right) =0, \crn &&
12m^{2}_{\chi^{\prime}}+3 \mu^{2}_{\chi} + g^{2} \left(
v^{2}_{\rho}-v^{\prime 2}_{\rho}-2v^{2}_{\chi}+2v^{\prime
2}_{\chi}\right)+ g^{\prime 2}\left( v^{2}_{\rho}-v^{\prime
2}_{\rho}-v^{2}_{\chi}+v^{\prime 2}_{\chi}\right) =0. \nn \eea

Let us consider the Higgs mass spectrum.

\subsection{Neutral scalar Higgs}

 Let us consider the mass spectrum of
 the neutral scalar Higgs bosons  in  the  model
  under consideration.
The mass Lagrangian of neutral scalar Higgs can be written  in the
form \bea \mathcal{L}_{\mathrm{H}}&=&-\frac{1}{2} \left(
H_{\rho},H_{\chi}, H_{\rho^{\prime}},H_{\chi^{\prime}} \right)
\mathcal{M}^2_{\mathrm{H}}~\left( H_{\rho},H_{\chi},
H_{\rho^{\prime}},H_{\chi^{\prime}} \right)^T,
\label{lrealmatrix1} \eea
 where
 \small{
 \bea
\mathcal{M}^2_{\mathrm{H}}  &=&
\frac{1}{3}(2g^2+g^{\prime2})v_{\chi'}^2
\label{NHiggsMatrix}\left(
\begin{array}{cccc}
t_1^2 & -a~t_1\tan\alpha & -a~ t_1t_2 &  a~ t_1 \\
-a~t_1\tan\alpha  &   \tan^2\alpha & a~ t_2 \tan\alpha & -\tan\alpha \\
-a~t_1& a~ t_2 \tan\alpha  & t_2^2 & -a~t_2 \\
a~ t_1 & -\tan\alpha  & -a~t_2 & 1 \\
\end{array}
\right),\crn &=& \frac{1}{3}(2g^2+g^{\prime2})v_{\chi'}^2
\mathcal{M}^2_{1\mathrm{H}}
 \eea
with \bea a &=& \frac{g^2+g^{\prime2}}{2g^2+g^{\prime2}},\;\;
(0<a<1),\crn t_1&=& \frac{v_{\rho}}{v_{\chi'}},\;\;\; t_2
=\frac{v_{\rho'}}{v_{\chi'}} \;\;\; (t_1,\;t_2 \ll
1)~~~\mathrm{and}\crn \tan\alpha&=& \frac{v_{\chi}}{v_{\chi'}}.
\eea Because $\det(\mathcal{M}^2_{1\mathrm{H}})=0$, we get a
zero-eigenvalue. It is convenient to diagonalize the neutral Higgs
mass matrices in two stages. First, we find the transformation for
original basis, particularly \bea
  H &=& C~H_1 \leftrightarrow \left(%
\begin{array}{c}
  H_{\rho} \\
   H_{\chi}\\
  H_{\rho'} \\
  H_{\chi'} \\
\end{array}
\right)= \left(
\begin{array}{cccc}
  0 & 1 & 0 & 0 \\
   \cos\alpha  & 0 & 0 & \sin\alpha \\
  0 & 0 & 1 & 0 \\
  \sin\alpha  & 0 & 0 & - \cos\alpha \\
\end{array}
\right)\left(
\begin{array}{c}
  H_{1\rho} \\
   H_{1\chi}\\
  H_{1\rho'} \\
  H_{1\chi'} \\
\end{array}
\right). \eea In the new basis $(H_{1\rho},\;
H_{1\chi},\;H_{1\rho'},\;H_{1\chi'})$,  we have \bea
M^2_{\mathrm{2H}}  &=& C^T  \mathcal{M}^2_{1\mathrm{H}} C\crn
&=& \left(%
\begin{array}{cccc}
0 & 0 &0&  0 \\
0& t_1^2  &  -a~t_1t_2 &- \frac{  a~ t_1}{\cos\alpha} \\
0& -a~t_1 t_2 & t_2^2 &  \frac{  a~ t_2}{\cos\alpha}  \\
0 & - \frac{  a~ t_1}{\cos\alpha}  &  \frac{  a~ t_2}{\cos\alpha}  &  \frac{  1}{\cos\alpha^2}  \\
\end{array}
\right)= \left(
   \begin{array}{cc}
     0 & 0 \\
     0 & M_{3\times 3}\\
   \end{array}
   \right).
\eea
We would like to remind the reader of the energy
  scale $v_\chi, v_{\chi^\prime} \gg v_{\rho},
 v_{\rho^\prime}$. This limit leads to $\tan \alpha \gg t_1,t_2$ and
 the matrix $M_3$ is a hierarchical matrix. Hence, it is very useful to use the method of
 block diagonalization in order to find the
eigenvectors and and eigenvalues of the matrix $M_3$.

 Let us rewrite matrix $\mathcal{M}^2_{\mathrm{2H}}$ in the basis
$(H_{1\chi'},\; H_{1\rho'},\; H_{1\chi}\;)$. In this basis, the matrix $M_3$ is
$3 \times 3$ matrix which has form as follows:
 \bea
  M_{3\times 3}= \left(
\begin{array}{ccc}
\frac{1}{\cos^2 \alpha } &  \frac{at_2}{\cos \alpha} & -\frac{at_1}{\cos \alpha} \\
\frac{at_2}{\cos \alpha} &   t_2^2 & -a~ t_1 t_2  \\
-\frac{at_1}{\cos \alpha}& -a~ t_1 t_2 & t_1^2  \\
  \end{array}
  \right).
 \eea
Next we can use an unitary matrix $U_1$ such as
 \begin{eqnarray}
U_1=\left(%
\begin{array}{ccc}
  1 & \frac{a t_2}{\cos\alpha} & -\frac{a t_1}{\cos \alpha} \\
 \frac{a t_2}{\cos\alpha} & 1 & 0 \\
 - \frac{a t_1}{\cos \alpha} & 0 & 1 \\
\end{array}%
\right)
\end{eqnarray}
in order to transform $M_3$ into the
approximately block-diagonal form and also we change the basis of
$H_1=(H_{1\chi},H_{1\rho},H_{1\rho'})$
into the new basis $H_2=(H_{2\chi},H_{2\rho},H_{2\rho'})$.
Details  are as follows
   \begin{eqnarray}
   H_2= U_1^{-1}H_1,
    \end{eqnarray}
 \begin{eqnarray}
   U_1^\dag M_3 U_1 \simeq
    \left(%
    \begin{array}{cc}
     \frac{1}{\cos^2\alpha} & 0 \\
      0 & M_{2\times 2} \\
    \end{array}%
    \right)
\label{huong1}\end{eqnarray}
with
\begin{eqnarray}
M_{2 \times 2}= \left(%
                \begin{array}{cc}
                  t_1^2-a^2t_2^2 & (a-1)at_1t_2 \\
                 (a-1)at_1t_2 & t_2^2-a^2t_1^2 \\
                \end{array}%
\right).
\end{eqnarray}
The Eq.(\ref{huong1}) proves the existence of the eigenvalues with
 value  $\tan^2 \alpha+1 $. The matrix $M_{2 \times 2}$ produces
two eigenvalues as follows
\bea
m_{3\rho} & = & \frac{1}{2}\left(
(1-a^2)(t_1^2+t_2^2)-\sqrt{(1+a^2)^2(t_1^2-t_2^2)^2-4(a-1)^2a^2t_1^2t_2^2}\right),\crn
m_{3\rho^\prime} & = &
\frac{1}{2}\left( (1-a^2)(t_1^2+t_2^2)+\sqrt{(1+a^2)^2(t_1^2-t_2^2)^2-4(a-1)^2a^2t_1^2t_2^2}\right)
\eea
with two eigenstates are, respectively
\bea
 H_{3\rho} &=& c_\zeta H_{2 \rho}-s_\zeta H_{2 \rho^\prime},\crn
 H_{3\rho^\prime} &= & s_\zeta H_{2 \rho}+c_\zeta H_{2 \rho^\prime}
\eea
with
$s_\zeta \equiv  \sin \zeta, c_\zeta\equiv \cos \zeta $ and $\zeta$
is determined through $\tan \zeta$,  as follows
\be
 \tan 2 \zeta=\frac{2a(1-a)t_1t_2}{(1+a)(t_1^2-t_2^2)}
\ee Let us summarize the neutral Higgs mass spectrum. There is one
massless Higgs namely $ \chi_{1}^\prime$ and there are
three massive states. One heavy Higgs is $(H_{2\chi})$ with mass
\begin{equation}
m_{H_{2\chi}^2 }=\frac{1}{3}(2g^2+g^{\prime 2})(1+ \tan^2 \alpha)
v_{\chi^\prime}^2.
\end{equation}
Two remaining Higgs are
$H_{3\rho},H_{3\rho^\prime}$
with  masses, respectively
\bea
m^2_{H_{3\rho}}&=&\frac{1}{3}(2g^2+g^{\prime 2})m_{3\rho}v_{\chi^\prime}^2,
\crn
m^2_{H_{3\rho^\prime}}&=&\frac{1}{3}(2g^2+g^{\prime 2})m_{3\rho^\prime}v_{\chi^\prime}^2.
\eea

\subsection{Pseudo-scalar Higgs}

The  model under  consideration contains four massless
pseudo-scalar Higgs bosons, namely $F_\rho,F_\chi, F_\rho^\prime,
F_\chi^\prime$, and  the mass matrix elements in this case are all
equal to zero. It means that all pseudoscalars are massless.

\subsection{Singly charged Higgs boson}

In the basis $\left( \rho^{-}, \rho^{\prime -}, \chi^{-},
\chi^{\prime -} \right)$,  the mass Lagrangian
 for  singly charged Higgs bosons has the form
\be \mathcal{L}^{\mathrm{Singly}}_{\mathrm{charged}}=\left(
\rho^{-}, \rho^{\prime -}, \chi^{-}, \chi^{\prime -}
\right)\mathcal{M}^{\mathrm{Singly}}_{\mathrm{charged}}\left(
\rho^{-}, \rho^{\prime -}, \chi^{-}, \chi^{\prime -} \right)^T \ee
with the mass matrix elements are  given by \bea
\mathcal{M}_{11}&=&\frac{g^{2}}{8}v^{2}_{\rho^{\prime}}, \,\
\mathcal{M}_{12}=- \frac{g^{2}}{8}v_{\rho}v_{\rho^{\prime}}, \,\
\mathcal{M}_{13}= \mathcal{M}_{14}= \mathcal{M}_{23}=
\mathcal{M}_{24}=0, \crn
\mathcal{M}_{22}&=&\frac{g^{2}}{8}v^{2}_{\rho},
\mathcal{M}_{33}=\frac{g^{2}}{8}v^{2}_{\chi^{\prime}},
\mathcal{M}_{44}=\frac{g^{2}}{8}v^{2}_{\chi},
\mathcal{M}_{34}=\frac{g^2}{8}v_\chi v_\chi^\prime.
\label{masssinglyhissm} \eea The matrix
$\mathcal{M}^{\mathrm{single}}_{\mathrm{charged}}$ produces two
massless states, namely \bea H_{\rho_1}^+&=&
\frac{1}{v_\rho^2+v_{\rho^\prime}^2}\left(v_\rho v_{\rho^\prime}
 \rho^{\prime+}+v_{\rho^\prime}^2 \rho^{+} \right),\\
 H_{\rho_2}^+&=& \frac{1}{v_\chi^2+v_{\chi^\prime}^2}\left(v_\chi v_{\chi^\prime}
 \chi^{\prime+}+v_{\chi^\prime}^2 \chi^{+} \right),
\eea
and two massive singly charged Higgs bosons
\bea
H_{\rho_3}^+&=& \frac{1}{v_\rho^2+v_{\rho^\prime}^2}\left(-v_\rho v_{\rho^\prime}
 \rho^{\prime+}+v_{\rho}^2 \rho^{+} \right),\\
 H_{\rho_4}^+&=& \frac{1}{v_\chi^2+v_{\chi^\prime}^2}\left(-v_\chi v_{\chi^\prime}
 \chi^{\prime+}+v_{\chi}^2 \chi^{+} \right)
\eea
and their eigenvalues are, respectively
\bea
m^2_{H_{\rho_3}^+}&=&\frac{g^2}{8}\left(
v^2_{\rho}+v^2_{\rho^\prime}\right)=m_W^2, \crn
m^2_{H_{\rho_4}^+}&=&\frac{g^2}{8}\left(
v^2_{\chi}+v^2_{\chi^\prime}\right)=m_V^2.
\label{comparreferre1}
\eea
The singly charged Higgs bosons part contains two massless states and two massive states. One
has mass equal to the mass of  the $W$ gauge boson  and other one has mass equal to those of the
 $V$ gauge boson.  This characteristic property  of the considered model is similar to that of
 the SUSY economical 3-3-1 model \cite{thuy}. We would like to
 to emphasized that  in SUSY models,  the Higgs self-couplings are the gauge couplings. Hence
 the Higgs mass spectrum can be related to the gauge mass spectrum.  One of the main points
  we would like to remind that because the Higgs sector
in SUSY economical 3-3-1 model and the model under consideration
 are very simple. Hence, we can
 easily  obtain the  Higgs mass spectrum. However the other SUSY versions of
  3-3-1 models, the Higgs sector
 is very complicated, and  it is hard to obtain the Higgs  mass spectrum.
 Hence, we cannot see the relation between the masses of the
  charged scalar and vector fields in other
  SUSY 3-3-1 versions.

\subsection{Doubly charged Higgs boson}

The model under consideration contains four doubly charged Higgs bosons,
namely $\rho^{--}$, $\chi^{--}$, $ \rho^{\prime --}$, $
\chi^{\prime --}$. On this basis, we obtain the mass
 matrix for doubly charged Higgs boson as follows
 \bea
     M_{H^{--}}^2  =\frac{g^2}{8}\left(%
      \begin{array}{cccc}
   t_2^2+t_3^2-1 & t_1t_3 & -t_1t_2 & -t_1 \\
   t_1 t_3 & t_1^2-t_2^2+1 & -t_2t_3 & -t_3 \\
   -t_1t_2 & -t_2t_3 & t_1^2-t_3^2+1 & t_2 \\
   -t_1 & -t_3 & t_2 & -t_1^2+t_2^2+t_3^2 \\
 \end{array}%
 \right).
  \label{huong2}\eea
  The mass matrix  in Eq.(\ref{huong2}) produces the mass
  eigenvalues
  \bea
   m^2_{H_1^{--}}&= & 0, \hs  m^2_{H_2^{--}}=
   \frac{g^2 }{8} (v^2_{\chi^\prime}
   -v^2_{\chi}+v^2_\rho-v^2_{\rho^\prime}),   \crn
   m^2_{H_3^{--}} &= & -m^2_{H_2^{--}},\,  m^2_{H_4^{--}} =
    \frac{g^2}{8} (v^2_{\chi^\prime}
    +v^2_{\chi}+v^2_\rho+v^2_{\rho^\prime})=m^2_{U^{--}}
  \label{huong3}\eea
  and their mass eigenvectors are, respectively
    \begin{eqnarray}
      H_1^{--} && =\frac{1}{\sqrt{1+t_1^2+t_2^2+t_3^2}}\left(-t_1
      \rho^{--}+t_3 \chi^{--}-t_2\rho^{\prime --}+  \chi^{\prime --}
      \right), \crn
      H_2^{--} && =
      \frac{1}{\sqrt{1+t_2^2}}\left(t_2 \chi^{--}+\rho^{\prime --} \right),
      \crn
      H_3^{--} && =
      \frac{1}{\sqrt{1+t_1^2}}\left(\rho^{--}+t_1 \chi^{\prime--}
       \right),
      \crn
      H_4^{--} && = C_{H_4^{--}}
      \left(-t_1 \rho^{--}-\frac{(1+t_1^2)t_3}{t_2^2+t_3^2}\chi^{--}
      +\frac{(1+t_1^2)t_2}{t_2^2+t_3^2}\rho^{\prime--}+\chi^{\prime --}\right)
\label{huong4}\eea
 with
$C_{H_4^{--}}=\sqrt{\frac{(1+t_1^2)(1+t_1^2+t_2^2+t_3^2)}{t_2^2+t_3^2}}.$

The mass spectrum of the doubly charged Higgs given in
Eqs. (\ref{huong3}) shows that the  model contains  one massive particle
 with mass equal to that of the  doubly charged
 gauge boson $U^{--}$  and at least one tachyon field, one
massless field $H_1^{--}$ which is identified to the
Goldstone boson. To remove tachyon in the model, we have
to include the following condition:
$v_{\chi^\prime}^2-v_\chi^2=v_\rho^2-v_{\rho^\prime}^2$. This
leads to appear two other massless particles $H_2^{--}, H_3^{--}$
in the doubly charged Higgs spectrum. The presence of these
particles maybe effect to the invisible Z bosons decay modes. Let
us consider the invisible decay modes of $Z^{\mu}$ into the
massless doubly charged Higgs, namely
 $Z^{\mu} \rightarrow H_{\rho^{--}_2} H_{ \rho^{++}_2}$,
$Z^{\mu} \rightarrow H_{\rho^{--}_3} H_{ \rho^{++}_3}$.  Fig.
\ref{invisible} predicts the invisible decay rate of $Z^{\mu}$
into the massless  doubly charged Higgs bosons by studying random
scan over the parameter space, such as $w= 10^3 \div 10^5
\mathrm{GeV}$, $t_v=\frac{v_\rho^\prime}{v_\rho} =0\div 100$,
$t_w=\frac{v_\chi^\prime}{v_\chi}
 = 0\div10$ and $v_\rho^\prime =246 $ GeV.   The obtained result predicts
the contribution of massless doubly charged Higgs into invisible
partial width  of Z decay modes is very suppressed. It is suitable
to limit on Z-decays into unknown new particles
 width $\Ga_{new} < 6.3  $ MeV  at 95\% confidence level given in
Ref.~\cite{invisible}. If we compare our predicted results with
constraint given in Ref.~\cite{invisible}, we obtain very hard
constraint on the $t_w$ parameter particularly $t_w = 0.65\div
0.85$. On the other hand,  Fig.\ref{invisible1} predicts the $Z$
bosons decay into two doubly charged Higgs decay width by studying
random scan over $t_w = 0.68 \div 0.8$, $v_\rho^\prime =246 $ GeV,
$w_\chi = 10^3 \div 10^5 $ GeV. The fig.\ref{invisible1} plays
probability to obtain the small invisible Z decay width
($\Gamma_{\mathrm{invisible}}\leq 2 $) MeV  is large and the
probability is almost independent upon parameter $t_v$. It
means that there is no constraint on the $t_v$ parameter in this
case.
\begin{figure}[h]
\begin{center}
\includegraphics[width=6.5cm,height=4cm]{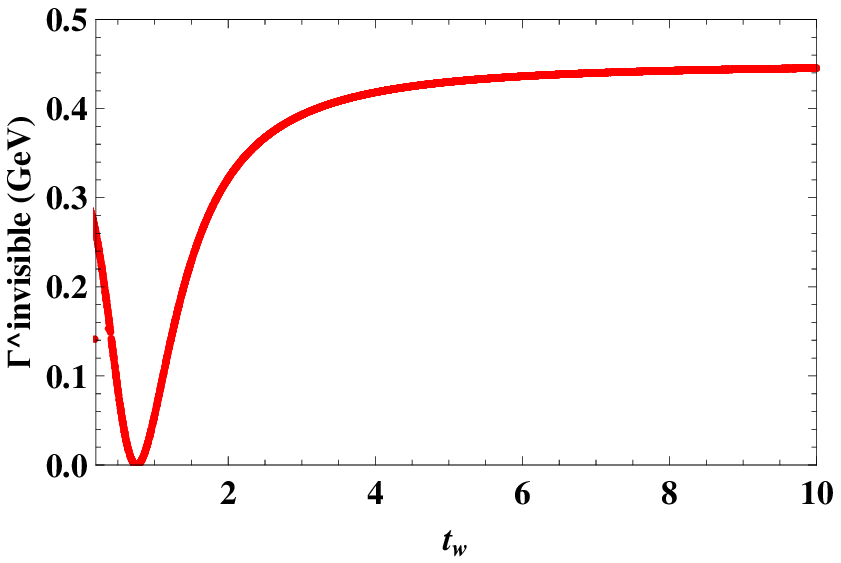}
\includegraphics[width=6.5cm,height=4cm]{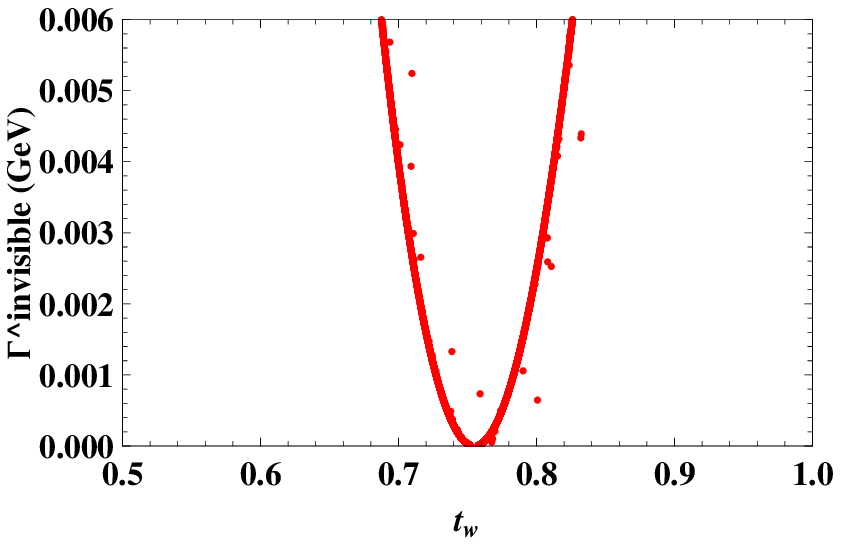}
\caption{\label{invisible}{The invisible decay rate of $Z^\mu$
into the massless of the doubly charged Higgs bosons by studying
random scan over the parameter space  $ w = 10^3 \div 10^5$ GeV,
$t_w = 0\div10$ and $t_v =0\div 100$, $v_\rho^\prime =246 $ GeV}.}
\end{center}
\end{figure}

\begin{figure}[h]
\begin{center}
\includegraphics[width=10cm,height=7cm]{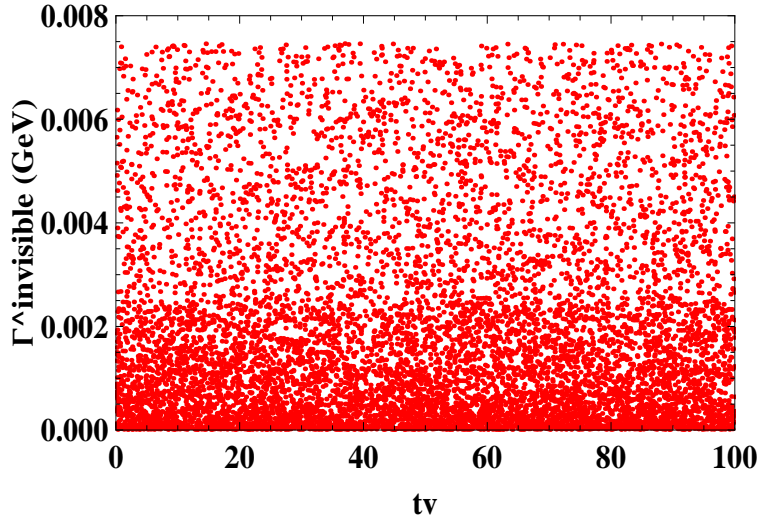}
\caption{\label{invisible1}{ The invisible decay rate of
$Z^\mu$ into the massless doubly charged Higgs bosons as the
function of $t_v$ by studying random scan over the parameter space
$w= 10^3 \div 10^5 $ GeV,$t_w = 0.68\div0.8$, $v_\rho^\prime =246
$ GeV}.}
\end{center}
\end{figure}

\newpage
\section{Conclusions}
\label{sec:con}

We have built  the supersymmetric version of the reduced
minimal 3-3-1 model with two Higgs triplets. We have  studied the
mass spectrum of all particles contained in the model. The
exact mass spectrum of gauge bosons is studied. In this sector
beyond the usual gauge bosons, $W^{\pm}, Z$ gauge bosons,  we have
two additional charged bosons, $V^{\pm}$ and $U^{\pm \pm}$,  and
one additional neutral gauge boson $Z^{\prime}$. The constraint on
the gauge mass is given by
$M_{Z^{\prime}}>M_{U}>M_{V}>M_{Z}>M_{W}$. In the charged-fermion
sector only the tau, top, bottom and charm quarks get their masses
at tree level, the others get their masses at one loop level. In
the neutrino sector only one neutrino gets mass at tree
level,  the others two $\nu_{\mu}$ and $\nu_{e}$ get their masses
at one loop level. The neutrino masses are smaller than those of
 the charged leptons.  It means that we explained  the
hierarchy  of fermion masses  in the model under consideration.
 In the Higgs sector, we can solve exactly the mass eigenstates and mass
eigenvalues for charged Higgs bosons. The masses of the massive
charged Higgs equal those of the charged gauge bosons, namely
$m^2_{H_{\rho_3}^\pm} =M^2_{W^\pm}, m^2_{H_{\rho_4}^\pm}
=M^2_{V^\pm}$  and $m^2_{H_4^{--}}=M^2_{U^{--}}$. In addition to
massive charged Higgs bosons, in the sector of  the doubly charged
Higgs bosons it  also appears  the tachyon field. If the tachyon
 field is removed, the model contains  two massless doubly
charged Higgs bosons.  By studying the effect of $Z\rightarrow
H_{2,3}^{++} H_{2,3}^{--}$ modes on invisible decay width of the
$Z$ bosons, we obtain the narrow constraint on $t_w = 0.65 \div
0.8$. In the neutral Higgs bosons, it is very hard to obtain the
exact  mass spectrum. However, with the help of the relation $
 u, u^\prime\ll w,w^\prime$, the diagonalization of neutral Higgs boson sector
has been performed by using the method of block diagonalization.
It leads to the neutral Higgs sector contained three massive
states and one massless particle. All pseudo-scalar particles are
massless.  Some of which are identified to the Goldstone bosons
and
 the remaining pseudo-scalar particles can be
identified to the axion. This analysis is not considered in details  in this paper.

\bc

{\it Acknowledgement}: \ec

 D. T. Huong would like to thank to P.
V.  Dong for very useful discussion on the invisible decay mode of
the  Z bosons. This research is funded by Vietnam  National
Foundation for Science and Technology Development (NAFOSTED) under
grant number 103.01-2011.63.

\appendix

\section{Lagrangian}
\label{sec:lagrangian}

We are going to write  the Lagrangians in terms of the fields in
this model

\subsection{Lepton Lagrangian}
\label{a1}
\begin{eqnarray}
\mathcal{L}_{\mathrm{Lepton}}&=&\int d^4\theta
     \hat{\bar{L}} \exp \left[ 2 \left( g \frac{\lambda^a}{2}\hat{V}^a \right)
\right] \hat{L} \nonumber \\
&=&\mathcal{L}^{\mathrm{lep}}_{llV}+
\mathcal{L}^{\mathrm{lep}}_{\tilde{l} \tilde{l}V}+
\mathcal{L}^{\mathrm{lep}}_{l \tilde{l} \tilde{V}}+
\mathcal{L}^{\mathrm{lep}}_{ \tilde{l} \tilde{l}V
V}+\mathcal{L}^{\mathrm{lep}}_{\mathrm{kin}}+\mathcal{L}^{\mathrm{lep}}_{\mathrm{F}}+
\mathcal{L}^{\mathrm{lep}}_{\mathrm{D}}, \label{lint}
\end{eqnarray}
The leptons in this model interact only with the weak
$\mathrm{SU(3)_L}$ boson, $V^a_{\mu}$,  and  they
 do not directly couple to the $\mathrm{U(1)}_X$
boson $V_{\mu}$.   The interaction between
 leptons and gauge bosons in component is given by
\begin{eqnarray}
\mathcal{L}^{\mathrm{lep}}_{llV}=\frac{g}{2}
\bar{L}\bar\sigma^{\mu}\lambda^a LV^a_{\mu}, \label{lepbos}
\end{eqnarray}
where $\lambda^a$ are the usual Gell-Mann matrices.The next part
is the slepton gauge boson interaction
\begin{eqnarray}
\mathcal{L}^{\mathrm{lep}}_{ \tilde{l} \tilde{l}V}=-
\frac{ig}{2}\left[
\tilde{L}\lambda^a\partial^{\mu}\bar{\tilde{L}}-
\bar{\tilde{L}}\lambda^a\partial^{\mu} \tilde{L}
\right]V^{a}_{\mu}.
\end{eqnarray}
The interaction between lepton-slepton-gaugino is given by the
following term
\begin{eqnarray}
\mathcal{L}^{\mathrm{lep}}_{l \tilde{l} \tilde{V}}=- \frac{ig}{
\sqrt{2}} ( \bar{L}\lambda^a\tilde{L}\bar{\lambda}^a_{A}-
\bar{\tilde{L}}\lambda^aL\lambda^a_{A}), \label{llchar}
\end{eqnarray}
and the four-interaction between sleptons and gauge bosons
\begin{eqnarray}
\mathcal{L}^{\mathrm{lep}}_{ \tilde{l} \tilde{l}VV}= \frac{g^2}{4}
V_{\mu}^aV^{b\mu} \bar{\tilde{L}} \lambda^{a}\lambda^{b}
\tilde{L}.
\end{eqnarray}
The kinetic parts of the leptons and sleptons are
\begin{eqnarray}
\mathcal{L}^{\mathrm{lep}}_{\mathrm{kin}}=- \vert \partial_{\mu}
\tilde{L}\vert ^2- iL \sigma^{\mu} \partial_{\mu} \bar{L}.
\end{eqnarray}
The last two terms in Eq.(\ref{lint}) are the usual $F$ and $D$
terms given by
\begin{eqnarray}
\mathcal{L}^{\mathrm{lep}}_{\mathrm{F}}&=& \vert F_{L} \vert^2,
\crn \mathcal{L}^{\mathrm{lep}}_{\mathrm{D}}&=&
\bar{\tilde{L}}\lambda^a\tilde{L} D^a.
\end{eqnarray}

\subsection{Quark  Lagrangian}
\label{a3}

\begin{eqnarray}
\mathcal{L}_{\mathrm{Quarks}}
     &=& \int d^{4}\theta\;\left[\,\hat{ \bar{Q}}_{1}
e^{2[g_{s}\hat{V}_{c}+g\hat{V}+(2g'/3)\hat{V}']} \hat{Q}_{1}
+\,\hat{\bar{Q}}_{\alpha}
e^{2[g_{s}\hat{V}_{c}+g\hat{\bar{V}}-(g'/3)\hat{V}']}
\hat{Q}_{\alpha}\right.\crn& +&\left. \hat{ \bar{u}}_{i}
e^{2[g_{s}\hat{\bar{V}}_{c}-(2g'/3)\hat{V}']} \hat{u}_{i} +\hat{
\bar{d}}_{i} e^{2[g_{s}\hat{\bar{V}}_{c}+(g'/3)\hat{V}']}
\hat{d}_{i}\,\right. \nonumber \\&+& \left.\,\hat{ \bar{J}}
e^{2[g_{s}\hat{\bar{V}}_{c} -(5g'/3)\hat{V}']} \hat{J} + \hat{
\bar{j}}_{\beta}
e^{2[g_{s}\hat{\bar{V}}_{c} +(4g'/3)\hat{V}']} \hat{j}_{\beta}\right] \nonumber \\
&=&\mathcal{L}_{qqV}+\mathcal{L}_{ \tilde{q}\tilde{q}V}+
\mathcal{L}_{q \tilde{q} \tilde{V}}+ \mathcal{L}_{
\tilde{q}\tilde{q}VV}
+\mathcal{L}^{\mathrm{quark}}_{\mathrm{kin}}+\mathcal{L}^{\mathrm{quark}}_{\mathrm{F}}+
\mathcal{L}^{\mathrm{quark}}_{\mathrm{D}}.
\end{eqnarray}
with $i,j,k=1,2,3$, $\alpha=2,3$ and $\beta=1,2$. In this case, as
in the lepton sector we can write
\begin{eqnarray}
\mathcal{L}^{\mathrm{quark}}_{\mathrm{kin}}&=&\tilde{Q}_i \square
\tilde{Q}^{*}_i+ \tilde{u}^{c}_i \square \tilde{u}^{c*}_i+
\tilde{d}^{c}_i \square \tilde{d}^{c*}_i+ \tilde{J}^{c}_i \square
\tilde{J}^{c*}_i- iQ_i \sigma^\mu
\partial_\mu \bar{Q_i}-
iu^{c}_{i} \sigma^\mu \partial_\mu \bar{u}^{c}_{i} \nonumber \\
&-&id^{c}_{i} \sigma^{\mu} \partial_{\mu} \bar{d}^{c}_{i}-
iJ^{c}_{i} \sigma^{\mu} \partial_{\mu} \bar{J}^{c}_{i}, \nonumber \\
\mathcal{L}^{\mathrm{quark}}_{\mathrm{F}}&=& \vert F_{Q_i} \vert^2
+
 \vert F_{u_i} \vert^2 + \vert F_{d_i} \vert^2 +
\vert F_{J_i} \vert^2, \nonumber \\
\mathcal{L}^{\mathrm{quark}}_{\mathrm{D}}&=& \frac{g_{s}}{2}(
\bar{\tilde{Q}}_{i}\lambda^a\tilde{Q}_{i} -
\bar{\tilde{u}}^{c}_{i}\lambda^{*a}\tilde{u}^{c}_{i}-
\bar{\tilde{d}}^{c}_{i}\lambda^{*a}\tilde{d}^{c}_{i}-
\bar{\tilde{J}}^{c}_{i}\lambda^{*a}\tilde{J}^{c}_{i})D^a_c \crn&+&
\frac{g}{2} \left( \bar{\tilde{Q}}_{3}\lambda^a\tilde{Q}_{3}-
\bar{\tilde{Q}}_{ \alpha}\lambda^{*a}\tilde{Q}_{ \alpha}  \right) D^{a} \nonumber \\
&+& \frac{g^{\prime}}{2\sqrt{6}} \left[ \frac{2}{3}
\bar{\tilde{Q}}_{3}\tilde{Q}_{3} -
 \frac{1}{3} \bar{\tilde{Q}}_{ \alpha}\tilde{Q}_{ \alpha}
- \frac{2}{3} \bar{\tilde{u}}^{c}_{i}\tilde{u}^{c}_{i}+
\frac{1}{3} \bar{\tilde{d}}^{c}_{i}\tilde{d}^{c}_{i}- \frac{5}{3}
\bar{\tilde{J}}^{c}\tilde{J}^{c}+ \frac{4}{3}
\bar{\tilde{j}}^{c}_{ \beta}\tilde{j}^{c}_{ \beta}
 \right] D, \nonumber \\
\mathcal{L}_{qqV}&=& \frac{g_{s}}{2} (
\bar{Q}_{i}\bar\sigma^{\mu}\lambda^a Q_{i}-
\bar{u}^{c}_i\bar\sigma^{\mu}\lambda^{*a} u^{c}_i-
\bar{d}^{c}_i\bar\sigma^{\mu}\lambda^{*a} d^{c}_i-
\bar{J}^{c}_i\bar\sigma^{\mu}\lambda^{*a} J^{c}_i)g^a_{\mu}  \nonumber \\
&+& \frac{g}{2}( \bar{Q}_{3}\bar\sigma^{\mu}\lambda^a Q_{3}-
\bar{Q}_{ \alpha}\bar\sigma^{\mu}\lambda^{*a} Q_{
\alpha})V^a_{\mu} \crn & + &\frac{g^{ \prime}}{2\sqrt{6}} \left(
\frac{2}{3} \bar{Q}_{3}\bar\sigma^{\mu} Q_{3}- \frac{1}{3}
\bar{Q}_{ \alpha}\bar\sigma^{\mu} Q_{ \alpha}- \frac{2}{3}
\bar{u}^{c}_i\bar\sigma^{\mu} u^{c}_i\right.\nonumber\\
&+&\left. \frac{1}{3} \bar{d}^{c}_i\bar\sigma^{\mu} d^{c}_i-
\frac{5}{3} \bar{J}^{c}\bar\sigma^{\mu} J^{c}+ \frac{4}{3}
\bar{j}^{c}_{ \beta}\bar\sigma^{\mu} j^{c}_{ \beta}
\right)B_{\mu}, \nonumber \\
\mathcal{L}_{ \tilde{q} \tilde{q}V}&=& \frac{-ig_{s}}{2} \left[
\left( \tilde{Q}_{i}\lambda^a\partial^{\mu}\bar{\tilde{Q}}_{i}-
\bar{\tilde{Q}}_{i}\lambda^a\partial^{\mu} \tilde{Q}_{i}-
\tilde{u}^{c}_i\lambda^{*a}\partial^{\mu}\bar{\tilde{u}}^{c}_i +
\bar{\tilde{u}}^{c}_i\lambda^{*a}\partial^{\mu} \tilde{u}^{c}_i \right.\right. \nonumber \\
&-& \left.\left.
\tilde{d}^{c}_i\lambda^{*a}\partial^{\mu}\bar{\tilde{d}}^{c}_i+
\bar{\tilde{d}}^{c}_i\lambda^{*a}\partial^{\mu} \tilde{d}^{c}_i -
\tilde{J}^{c}_i\lambda^{*a}\partial^{\mu}\bar{\tilde{J}}^{c}_i+
\bar{\tilde{J}}^{c}_i\lambda^{*a}\partial^{\mu} \tilde{J}^{c}_i\right)g^a_{\mu} \right] \nonumber \\
&-& \frac{i g}{2} \left(
\tilde{Q}_{3}\lambda^a\partial^{\mu}\bar{\tilde{Q}}_{3}-
\bar{\tilde{Q}}_{3}\lambda^a\partial^{\mu} \tilde{Q}_{3}-
\tilde{Q}_{ \alpha}\lambda^{*a}\partial^{\mu}\bar{\tilde{Q}}_{
\alpha}+
\bar{\tilde{Q}}_{ \alpha}\lambda^{*a}\partial^{\mu} \tilde{Q}_{ \alpha}\right)V^a_{\mu} \nonumber \\
&-& \frac{ig^{ \prime}}{2\sqrt{6}} \left[ \frac{2}{3} (
\tilde{Q}_{3}
\partial^{\mu}\bar{\tilde{Q}}_{3}- \bar{\tilde{Q}}_{3} \partial^{\mu}
\tilde{Q}_{3})- \frac{1}{3} ( \tilde{Q}_{ \alpha}
\partial^{\mu}\bar{\tilde{Q}}_{ \alpha}-
\bar{\tilde{Q}}_{ \alpha} \partial^{\mu} \tilde{Q}_{
\alpha})\right.\crn&-&\left. \frac{2}{3} ( \tilde{u}^{c}_i
\partial^{\mu}\bar{\tilde{u}}^{c}_i-
\bar{\tilde{u}}^{c}_i \partial^{\mu} \tilde{u}^{c}_i) +
\frac{1}{3} (\tilde{d}^{c}_i \partial^{\mu}\bar{\tilde{d}}^{c}_i-
\bar{\tilde{d}}^{c}_i
\partial^{\mu} \tilde{d}^{c}_i)\right.\crn&-&\left. \frac{5}{3} (\tilde{J}^{c}
\partial^{\mu}\bar{\tilde{J}}^{c}- \bar{\tilde{J}}^{c} \partial^{\mu}
\tilde{J}^{c})+ \frac{4}{3} (\tilde{j}^{c}_{ \beta}
\partial^{\mu}\bar{\tilde{j}}^{c}_{ \beta}-
\bar{\tilde{j}}^{c}_{ \beta} \partial^{\mu} \tilde{j}^{c}_{ \beta}) \right] B_{\mu}, \nonumber \\
\mathcal{L}_{q \tilde{q} \tilde{V}}&=& \frac{-ig_{s}}{ \sqrt{2}}
\left[ ( \bar{Q}_{i}\lambda^a\tilde{Q}_{i}-
\bar{u}^{c}_i\lambda^{*a}\tilde{u}^{c}_i-
\bar{d}^{c}_i\lambda^{*a}\tilde{d}^{c}_i-
\bar{J}^{c}_i\lambda^{*a}\tilde{J}^{c}_i) \bar{\lambda}^a_{c} \right. \nonumber \\
&-& \left. ( \bar{\tilde{Q}}_{i}\lambda^aQ_{i}-
\bar{\tilde{u}}^{c}_i\lambda^{*a}u^{c}_i-
\bar{\tilde{d}}^{c}_i\lambda^{*a}d^{c}_i-
\bar{\tilde{J}}^{c}_i\lambda^{*a}J^{c}_i)
\lambda^a_{c} \right] \nonumber \\
&-& \frac{i g}{ \sqrt{2}} \left[ (
\bar{Q}_{3}\lambda^a\tilde{Q}_{3}-
\bar{Q}_{\alpha}\lambda^{*a}\tilde{Q}_{\alpha})\bar{\lambda}^a_{A}-
( \bar{\tilde{Q}}_{3}\lambda^aQ_{3}-
\bar{\tilde{Q}}_{\alpha}\lambda^{*a}Q_{\alpha}) \lambda^a_{A} \right] \nonumber \\
&-& \frac{i g'}{2\sqrt{3}} \left[ \left( \frac{2}{3}
\bar{Q}_{3}\tilde{Q}_{3}-
\frac{1}{3}\bar{Q}_{\alpha}\tilde{Q}_{\alpha}-\frac{2}{3}
\bar{u}^{c}_i\tilde{u}^{c}_i+ \frac{1}{3}
\bar{d}^{c}_i\tilde{d}^{c}_i-\frac{5}{3} \bar{J}^{c}\tilde{J}^{c}+
\frac{4}{3} \bar{j}^{c}_{ \beta}\tilde{j}^{c}_{ \beta} \right)
\bar{\lambda}_{B}
\right. \nonumber \\
&-& \left. \left( \frac{2}{3}\bar{\tilde{Q}}_{3}Q_{3}-
\frac{1}{3}\bar{\tilde{Q}}_{\alpha}Q_{\alpha}- \frac{2}{3}
\bar{\tilde{u}}^{c}_iu^{c}_i+\frac{1}{3}
\bar{\tilde{d}}^{c}_id^{c}_i- \frac{5}{3}
\bar{\tilde{J}}^{c}J^{c}+ \frac{4}{3} \bar{\tilde{j}}^{c}_{
\beta}j^{c}_{ \beta} \right) \lambda_B \right],
\nonumber \\
\mathcal{L}_{ \tilde{q} \tilde{q}VV}&=& \frac{-1}{4} \left[
g_{s}^2( \bar{\tilde{Q}}_{i}\lambda^{a}\lambda^{b} \tilde{Q}_{i}+
\bar{\tilde{u}}^{c}_i\lambda^{*a}\lambda^{*b}
\tilde{u}^{c}_i\right.\crn&+&\left.
\bar{\tilde{d}}^{c}_i\lambda^{*a}\lambda^{*b} \tilde{d}^{c}_i+
\bar{\tilde{J}}^{c}_i\lambda^{*a}\lambda^{*b}
\tilde{J}^{c}_i)g^a_\mu g^{b\mu}
\right] \nonumber \\
&-& \frac{1}{4} \left[ g^2(
\bar{\tilde{Q}}_{3}\lambda^{a}\lambda^{b} \tilde{Q}_{3}+
\bar{\tilde{Q}}_{ \alpha}\lambda^{*a}\lambda^{*b} \tilde{Q}_{
\alpha}) \right] V^a_\mu V^{b\mu}\crn&-&  \frac{1}{2} \left[
g_{s}g( \bar{\tilde{Q}}_{3}\lambda^{a}\lambda^{b} \tilde{Q}_{3}+
\bar{\tilde{Q}}_{ \alpha}\lambda^{a}\lambda^{*b} \tilde{Q}_{
\alpha})
\right] g^a_\mu V^{b\mu} \nonumber \\
&-&  \frac{g_{s}g^{ \prime}}{2\sqrt{6}} \left[ \frac{2}{3}
\bar{\tilde{Q}}_{3}\lambda^a \tilde{Q}_{3}- \frac{1}{3}
\bar{\tilde{Q}}_{ \alpha}\lambda^a \tilde{Q}_{ \alpha}+
\frac{2}{3} \bar{\tilde{u}}^{c}_i\lambda^{*a}
\tilde{u}^{c}_i\right.\crn&-&\left. \frac{1}{3}
\bar{\tilde{d}}^{c}_i\lambda^{*a} \tilde{d}^{c}_i+ \frac{5}{3}
\bar{\tilde{J}}^{c}\lambda^{*a} \tilde{J}^{c} -\frac{4}{3}
\bar{\tilde{j}}^{c}_{ \beta}\lambda^{*a} \tilde{j}^{c}_{ \beta}
\right] g^{a\mu}B_{\mu} \nonumber\\&-& \frac{gg^{ \prime}
}{2\sqrt{6}}\left[ \frac{2}{3} \bar{\tilde{Q}}_{3}\lambda^a
\tilde{Q}_{3}+ \frac{1}{3} \bar{\tilde{Q}}_{ \alpha}\lambda^{*a}
\tilde{Q}_{ \alpha} \right] V^{a\mu}B_{\mu} \crn&-& \frac{g^{
\prime 2}}{24}  \left[
\frac{4}{9}(\bar{\tilde{Q}}_{3}\tilde{Q}_{3}+\bar{\tilde{u}}^{c}_i\tilde{u}^{c}_i)+
\frac{1}{9}(\bar{\tilde{Q}}_{\alpha}\tilde{Q}_{\alpha}+
\bar{\tilde{d}}^{c}_i\tilde{d}^{c}_i)\right.\nonumber\\&+&\left.
\frac{25}{9} \bar{\tilde{J}}^{c}\tilde{J}^{c}+ \frac{16}{9}
\bar{\tilde{j}}^{c}_{\beta}\tilde{j}^{c}_{\beta} \right] B^{\mu}
B_{\mu},
\end{eqnarray}
where we used the following short identities $J_{1}=J$,
$J_{2}=j_{1}$ and $J_{3}=j_{2}$.

\subsection{Scalar Lagrangian}
\label{a2}

\begin{eqnarray}
 \mathcal{L}_{\mathrm{Scalar}}
     &=&  \int d^{4}\theta\;\left[\,~
     \hat{ \bar{ \rho}}e^{2g\hat{V}+g'\hat{V}'}\hat{ \rho} +
\hat{ \bar{
\chi}}e^{2g\hat{V}-g'\hat{V}'}\hat{\chi}\right.\crn&+&\left.
\hat{\bar{ \rho}}^\prime e^{2g\hat{\bar V}-g'\hat{\bar V}'} \hat{
\rho}^\prime + \hat{ \bar{ \chi}}^\prime e^{2g\hat{\bar
V}+g'\hat{\bar V}'}
\hat{ \chi}^\prime \right] \nonumber \\
&=&\mathcal{L}^{\mathrm{scalar}}_{\mathrm{F}}+\mathcal{L}^{\mathrm{scalar}}_{\mathrm{D}}+
\mathcal{L}_{\mathrm{Higgs}}+\mathcal{L}_{\mathrm{Higgsinos}}+\mathcal{L}_{H
\tilde{H} \tilde{V}}, \label{hint}
\end{eqnarray}
where the terms with the auxiliary fields can be rewritten as
\begin{eqnarray}
\mathcal{L}^{\mathrm{scalar}}_{\mathrm{F}}&=& \vert F_{\rho}
\vert^2+ \vert F_{\chi} \vert^2 + \vert F_{\rho^{\prime}} \vert^2
+
\vert F_{\chi^{\prime}} \vert^2 , \nonumber \\
\mathcal{L}^{\mathrm{scalar}}_{\mathrm{D}}&=& \frac{g}{2} \left[
\bar{\rho}\lambda^a\rho+ \bar{\chi}\lambda^a\chi-
\bar{\rho}^{\prime}\lambda^{* a}\rho^{\prime}-
\bar{\chi}^{\prime}\lambda^{* a}\chi^{\prime} \right] D^{a}
\crn&+& \frac{g^{ \prime}}{2\sqrt{6}} \left[ \bar{\rho}\rho-
\bar{\chi}\chi- \bar{\rho}^{\prime}\rho^{\prime}+
\bar{\chi}^{\prime}\chi^{\prime} \right]D,
\label{scalarauxiliarfields}
\end{eqnarray}
while the kinetics terms are
\begin{eqnarray}
\mathcal{L}_{\mathrm{Higgs}}&=& (\mathcal{D}_{\mu}
\rho)^{\dagger}(\mathcal{D}^{\mu} \rho)+ (\mathcal{D}_{\mu}
\chi)^{\dagger}(\mathcal{D}^{\mu} \chi)\crn&+&
(\overline{\mathcal{D}_{\mu}}
\rho^{\prime})^{\dagger}(\overline{\mathcal{D}^{\mu}}
\rho^{\prime})+ (\overline{\mathcal{D}_{\mu}}
\chi^{\prime})^{\dagger}(\overline{\mathcal{D}^{\mu}}
\chi^{\prime}),
\nonumber \\
\mathcal{L}_{\mathrm{Higgsinos}}&=& i \bar{\hat{\rho}}
\bar{\sigma}^{\mu}\mathcal{D}_{\mu}\hat{\rho}+ i \bar{\hat{\chi}}
\bar{\sigma}^{\mu}\mathcal{D}_{\mu}\hat{\chi}+ i
\bar{\hat{\rho}}^{\prime}
\bar{\sigma}^{\mu}\overline{\mathcal{D}_{\mu}}\hat{\rho}^{\prime}+
i \bar{\hat{\chi}}^{\prime}
\bar{\sigma}^{\mu}\overline{\mathcal{D}_{\mu}}\hat{\chi}^{\prime}.
\end{eqnarray}
The covariant derivatives are given by
\begin{eqnarray}
\mathcal{D}_{\mu}\phi_{i}  &=&  \partial_{\mu}\phi_{i} - ig\left(
\vec{V}_{\mu} .\frac{\vec{\lambda}}{2}\right)^{j}_{i}\phi_{j} -
ig^\prime X_{\phi_{i}}T^9B_{\mu}\phi_{i}, \nonumber \\
\overline{\mathcal{D}_{\mu}}\phi_{i}  &=&  \partial_{\mu}\phi_{i}
- ig\left( \vec{V}_{\mu}
.\frac{\vec{\overline{\lambda}}}{2}\right)^{j}_{i}\phi_{j} -
ig^\prime X_{\phi_{i}}T^9B_{\mu}\phi_{i}. \label{dct1}
\end{eqnarray}
The interaction between the scalar-gaugino-higgsino is given by
\begin{eqnarray}
\mathcal{L}_{H \tilde{H} \tilde{V}}&=&- \frac{ig}{ \sqrt{2}}
\left[ \bar{\tilde{\rho}}\lambda^a\rho\bar{\lambda}^a_{A} -
\bar{\rho}\lambda^a\tilde{\rho}\lambda^a_{A}+
\bar{\tilde{\chi}}\lambda^a\chi\bar{\lambda}^a_{A} -
\bar{\chi}\lambda^a\tilde{\chi}\lambda^a_{A}\right. \nonumber \\
&-& \left. \bar{\tilde{\rho}}^{\prime}\lambda^{*
a}\rho^{\prime}\bar{\lambda}^a_{A} + \bar{\rho}^{\prime}\lambda^{*
a}\tilde{\rho}^{\prime}\lambda^a_{A}
-\bar{\tilde{\chi}}^{\prime}\lambda^{*
a}\chi^{\prime}\bar{\lambda}^a_{A} + \bar{\chi}^{\prime}\lambda^{*
a}\tilde{\chi}^{\prime}\lambda^a_{A} \right]  \nonumber \\
&-& \frac{ig^{ \prime}}{2 \sqrt{3}} \left[
\bar{\tilde{\rho}}\rho\bar{\lambda}_{B}
-\bar{\rho}\tilde{\rho}\lambda_{B}-
\bar{\tilde{\chi}}\chi\bar{\lambda}_{B} \right.\crn&+&\left.
\bar{\chi}\tilde{\chi}\lambda_{B} -
\bar{\tilde{\rho}}^{\prime}\rho^{\prime}\bar{\lambda}_{B}
+\bar{\rho}^{\prime}\tilde{\rho}^{\prime}\lambda_{B} +
\bar{\tilde{\chi}}^{\prime}\chi^{\prime}\bar{\lambda}_{B}
-\bar{\chi}^{\prime}\tilde{\chi}^{\prime}\lambda_{B} \right],
\label{mix1}
\end{eqnarray}

\subsection{Gauge Lagrangian}
\label{a3}

\begin{eqnarray}
\mathcal{L}_{\mathrm{Gauge}}&=& \frac{1}{4} \left[ \int
d^{2}\theta\;\left ( {W}^{a}_{c}{W}^{a}_{c}+{W}^{a}{W}^{a}+{W}^{
\prime}{W}^{ \prime}\right)\right.\crn&+&\left. \int
d^{2}\bar{\theta}\; \left(
\bar{W}^{a}_{c}\bar{W}^{a}_{c}+\bar{W}^{a}\bar{W}^{a}+
\bar{W}^{ \prime}\bar{W}^{ \prime}\right)\,\right] \nonumber \\
&=&\mathcal{L}_{\mathrm{dc}}+\mathcal{L}^{\mathrm{gauge}}_{D}.
\label{lgauge}
\end{eqnarray}
The kinetic term has the following form
\begin{eqnarray}
\mathcal{L}_{\mathrm{dc}}&=&- \frac{1}{4} \left(
G^{a\mu\nu}G^{a}_{\mu\nu}+W^{a\mu\nu}W^{a}_{\mu\nu}+B^{\mu\nu}B_{\mu\nu}
\right) \crn&-&i \left( \bar{\lambda}^{a}_{C}
\bar{\sigma}^{\mu}\mathcal{D}^{C}_{\mu}\lambda^{a}_{C}+
\bar{\lambda}^{a}_{A}
\bar{\sigma}^{\mu}\mathcal{D}^{L}_{\mu}\lambda^{a}_{A}+
\bar{\lambda}_{B}\bar{\sigma}^{\mu}\partial_{\mu}\lambda_{B} \right) \,\ , \nonumber \\
\label{tutty}
\end{eqnarray}
with
\begin{eqnarray}
G^{a}_{\mu\nu}&=&
\partial_{\mu}g^{a}_{\nu}-\partial_{\nu}g^{a}_{\mu}-gf^{abc}g^{b}_{\mu}
g^{c}_{\nu}, \nonumber \\
W^{a}_{\mu\nu}&=&
\partial_{\mu}V^{a}_{\nu}-\partial_{\nu}V^{a}_{\mu}-gf^{abc}V^{b}_{\mu}
V^{c}_{\nu}, \nonumber \\
B_{\mu\nu}&=& \partial_{\mu}B_{\nu}-\partial_{\nu}B_{\mu}, \nonumber \\
\mathcal{D}^{C}_{\mu}\lambda^{a}_{C}&=&\partial_{\mu}\lambda^{a}_{C}-
g_{s}f^{abc}g^{b}_{\mu}\lambda^{c}_{C}, \crn
\mathcal{D}^{L}_{\mu}\lambda^{a}_{A}&=&\partial_{\mu}\lambda^{a}_{A}-gf^{abc}V^{b}_{\mu}\lambda^{c}_{A},
\label{gauge2}
\end{eqnarray}
where $f^{abc}$ are the structure constants of the gauge group
$\mathrm{SU(3)}$, and we have the usual self-interactions (cubic
and quartic) of the gauge bosons.
 The last term in Eq.(\ref{lgauge}) is
\begin{eqnarray}
\mathcal{L}^{\mathrm{gauge}}_{\mathrm{D}}=\frac{1}{2}\left(
D^{a}_{C}D^{a}_{C}+D^{a}D^{a}+DD \right). \label{gaugedfields}
\end{eqnarray}

\subsection{Superpotential}
\label{a4}

The superpotential of the model is given in  Eq.(\ref{sp1}).  The
superpotential in terms of the fields is given by
\begin{eqnarray}
W_{2}&=&\mathcal{L}^{W2}_{F}+\mathcal{L}_{\mathrm{HMT}}, \nonumber \\
W_{3}&=&\mathcal{L}^{W3}_{F}+\mathcal{L}_{ll
\tilde{l}}+\mathcal{L}_{llH}+ \mathcal{L}_{l \tilde{l}
\tilde{H}}+\mathcal{L}_{l\tilde{H}H}\crn&+&
\mathcal{L}_{\tilde{l}H H}+\mathcal{L}_{qqH}+\mathcal{L}_{q
\tilde{q} \tilde{H}}+ \mathcal{L}_{lq
\tilde{q}}+\mathcal{L}_{\tilde{l}q \tilde{q}}, \nonumber \\
\end{eqnarray} where
\begin{eqnarray}
\mathcal{L}^{W2}_{F}&=& \frac{\mu_{ \rho}}{2}( \rho
F_{\rho^{\prime}}+ \rho^{\prime} F_{ \rho}) + \frac{\mu_{
\chi}}{2}( \chi F_{\chi^{\prime}}+ \chi^{\prime} F_{ \chi}),
\nonumber \\
\mathcal{L}_{\mathrm{HMT}}&=&- \frac{\mu_{ \rho}}{2} \tilde{
\rho}_i \tilde{ \rho}^{\prime}_i-
\frac{\mu_{ \chi}}{2} \tilde{ \chi}_i \tilde{ \chi}^{\prime}_i, \nonumber \\
\mathcal{L}_{F}^{W_3}&=& \frac{1}{3}\left.[ 3 \lambda_{1} \epsilon
F_L \tilde{L} \tilde{L}+\lambda_{2} \epsilon (F_{L}\chi \rho+
\tilde{L}F_{\chi}\rho+ \tilde{L} \chi F_{\rho})
\right.\crn&+&\left.
\kappa_{1}(F_{Q_1}\rho^{\prime}\tilde{d}^{c}_{i}+\tilde{Q}_1F_{\rho^{\prime}}
\tilde{d}^{c}_{i}+\tilde{Q}_1\rho^{\prime}F_{d_{i}})\right. \nonumber \\
&+& \left.\kappa_{2}(F_{Q_1}\chi^{\prime}\tilde{J}^{c}+
\tilde{Q}_1F_{\chi^{\prime}}\tilde{J}^{c}+\tilde{Q}_1\chi^{\prime}F_{J})
\right.\crn&+&\left.
\kappa_{3}(F_{Q_{\alpha}}\rho\tilde{u}^{c}_{i}+
\tilde{Q}_{\alpha}F_{\rho}\tilde{u}^{c}_{i}+\tilde{Q}_{\alpha}\rho
F_{u_{i}})\right. \nonumber \\ &+&\left.
\kappa_{4}(F_{Q_{\alpha}}\chi \tilde{j}^{c}_{\beta}+
\tilde{Q}_{\alpha}F_{\chi}\tilde{j}^{c}_{\beta}+\tilde{Q}_{\alpha}\chi
F_{j_{\beta}}) \right.\crn&+&\left.
\kappa_{5}(F_{Q_{\alpha}}\tilde{L}
\tilde{d}^{c}_{i}+\hat{Q}_{\alpha}F_{L}\tilde{d}^{c}_{i}+
\tilde{Q}_{\alpha}\tilde{L}F_{d_{i}})\right], \nonumber \\
\mathcal{L}_{ll \tilde{l}}&=&- \frac{ \lambda_{1}}{3} \epsilon
(LL\tilde{L}+\tilde{L}LL+L\tilde{L}L),\nonumber\\
\mathcal{L}_{l\tilde{H}H}&=&- \frac{ \lambda_{2}}{3} \epsilon
(L \tilde{\chi} \rho+L \chi \tilde{\rho}), \nonumber \\
\mathcal{L}_{qqH}&=&- \frac{1}{3}[ \kappa_{1}
Q_1\rho^{\prime}d^{c}_{i}+ \kappa_{2} Q_1\chi^{\prime}J^{c}+
\kappa_{3} Q_{\alpha}\rho u^{c}_{i}+
\kappa_{4} Q_{\alpha}\chi j^{c}_{\beta}], \nonumber \\
\mathcal{L}_{q \tilde{q} \tilde{H}}&=&- \frac{1}{3}[
\kappa_{1}(Q_1 \tilde{d}^{c}_{i}+ \tilde{Q}_1d^{c}_{i}) \tilde{
\rho}^{\prime}+ \kappa_{2}(Q_1 \tilde{J}^{c}+ \tilde{Q}_1J^{c})
\tilde{ \chi}^{\prime}
\nonumber \\
&+& \kappa_{3}(Q_{\alpha} \tilde{u}^{c}_{i}+
\tilde{Q}_{\alpha}u^{c}_{i}) \tilde{ \rho}+ \kappa_{4}(Q_{\alpha}
\tilde{j}^{c}_{\beta}+ \tilde{Q}_{\alpha}j^{c}_{\beta})
\tilde{ \chi}], \nonumber \\
\mathcal{L}_{lq \tilde{q}}&=&- \frac{
\kappa_{5}}{3}(Q_{\alpha}\tilde{d}^{c}_{i}+
\tilde{Q}_{\alpha}d^{c}_{i})L,\nonumber\\
\mathcal{L}_{ \tilde{l}qq}&=&- \frac{ \kappa_{5}}{3} Q_{ \alpha}
\tilde{L}
d^{c}_{i}, \nonumber\\
\mathcal{L}_{\tilde{l}HH}&=&- \frac{ \lambda_{2}}{3} \tilde{L}
\chi \rho . \label{HMT}
\end{eqnarray}

\subsection{Superpotential}
\label{a4}

The superpotential of the model is given in  Eq.(\ref{sp1}).  The
superpotential in terms of the fields are given by
\begin{eqnarray}
W_{2}&=&\mathcal{L}^{W2}_{F}+\mathcal{L}_{ \hat{ \eta}L}+
\mathcal{L}_{\mathrm{HMT}}, \nonumber \\
W_{3}&=&\mathcal{L}^{W3}_{F}+\mathcal{L}_{ll
\hat{l}}+\mathcal{L}_{llH}+ \mathcal{L}_{l \hat{l}
\hat{H}}+\mathcal{L}_{l\hat{H}H}\nonumber\\&+&
\mathcal{L}_{\hat{l}H H}+\mathcal{L}_{qqH}+\mathcal{L}_{q \hat{q}
\hat{H}}+ \mathcal{L}_{lq \hat{q}}+\mathcal{L}_{\hat{l}q \hat{q}},
\nonumber \\ \end{eqnarray} where
\begin{eqnarray}
\mathcal{L}^{W2}_{F}&=& \frac{\mu_{ \rho}}{2}( \rho
F_{\rho^{\prime}}+ \rho^{\prime} F_{ \rho}) + \frac{\mu_{
\chi}}{2}( \chi F_{\chi^{\prime}}+ \chi^{\prime} F_{ \chi}),
\nonumber \\
\mathcal{L}_{\mathrm{HMT}}&=&- \frac{\mu_{ \rho}}{2} \hat{ \rho}_i
\hat{ \rho}^{\prime}_i-
\frac{\mu_{ \chi}}{2} \hat{ \chi}_i \hat{ \chi}^{\prime}_i, \nonumber \\
\mathcal{L}_{F}^{W_3}&=& \frac{1}{3}[ 3 \lambda_{1} \epsilon F_L
\hat{L} \hat{L}+\lambda_{2} \epsilon (F_{L}\chi \rho+
\hat{L}F_{\chi}\rho+ \hat{L} \chi F_{\rho}) \nonumber \\ &+&
\kappa_{1}(F_{Q_1}\rho^{\prime}\hat{d}^{c}_{i}+\hat{Q}_1F_{\rho^{\prime}}
\hat{d}^{c}_{i}+\hat{Q}_1\rho^{\prime}F_{d_{i}}) \nonumber \\ &+&
\kappa_{2}(F_{Q_1}\chi^{\prime}\hat{J}^{c}+
\hat{Q}_1F_{\chi^{\prime}}\hat{J}^{c}+\hat{Q}_1\chi^{\prime}F_{J})
\nonumber \\
&+& \kappa_{3}(F_{Q_{\alpha}}\rho\hat{u}^{c}_{i}+
\hat{Q}_{\alpha}F_{\rho}\hat{u}^{c}_{i}+\hat{Q}_{\alpha}\rho
F_{u_{i}}) \nonumber \\ &+& \kappa_{4}(F_{Q_{\alpha}}\chi
\hat{j}^{c}_{\beta}+
\hat{Q}_{\alpha}F_{\chi}\hat{j}^{c}_{\beta}+\hat{Q}_{\alpha}\chi
F_{j_{\beta}}) \nonumber \\
&+& \kappa_{5}(F_{Q_{\alpha}}\hat{L}
\hat{d}^{c}_{i}+\hat{Q}_{\alpha}F_{L}\hat{d}^{c}_{i}+
\hat{Q}_{\alpha}\hat{L}F_{d_{i}})], \nonumber \\
\mathcal{L}_{ll \hat{l}}&=&- \frac{ \lambda_{1}}{3} \epsilon
(LL\hat{L}+\hat{L}LL+L\hat{L}L),\nonumber\\
\mathcal{L}_{l\hat{H}H}&=&- \frac{ \lambda_{2}}{3} \epsilon
(L \hat{\chi} \rho+L \chi \hat{\rho}), \nonumber \\
\mathcal{L}_{qqH}&=&- \frac{1}{3}[ \kappa_{1}
Q_1\rho^{\prime}d^{c}_{i}+ \kappa_{2} Q_1\chi^{\prime}J^{c}+
\kappa_{3} Q_{\alpha}\rho u^{c}_{i}+
\kappa_{4} Q_{\alpha}\chi j^{c}_{\beta}], \nonumber \\
\mathcal{L}_{q \hat{q} \hat{H}}&=&- \frac{1}{3}[ \kappa_{1}(Q_1
\hat{d}^{c}_{i}+ \hat{Q}_1d^{c}_{i}) \hat{ \rho}^{\prime}+
\kappa_{2}(Q_1 \hat{J}^{c}+ \hat{Q}_1J^{c}) \hat{ \chi}^{\prime}
\nonumber \\
&+& \kappa_{3}(Q_{\alpha} \hat{u}^{c}_{i}+
\hat{Q}_{\alpha}u^{c}_{i}) \hat{ \rho}+ \kappa_{4}(Q_{\alpha}
\hat{j}^{c}_{\beta}+ \hat{Q}_{\alpha}j^{c}_{\beta})
\hat{ \chi}], \nonumber \\
\mathcal{L}_{lq \hat{q}}&=&- \frac{
\kappa_{5}}{3}(Q_{\alpha}\hat{d}^{c}_{i}+
\hat{Q}_{\alpha}d^{c}_{i})L,\nonumber\\
\mathcal{L}_{ \hat{l}qq}&=&- \frac{ \kappa_{5}}{3} Q_{ \alpha}
\hat{L}
d^{c}_{i}, \nonumber\\
\mathcal{L}_{\hat{l}HH}&=&- \frac{ \lambda_{2}}{3} \hat{L} \chi
\rho . \label{HMT}
\end{eqnarray}

\end{document}